\documentclass{article}
\usepackage[numbers]{natbib}
\usepackage[preprint]{neurips_2024}

\usepackage{graphicx}
\usepackage{multicol,multirow}
\usepackage{amsmath,amssymb,amsfonts}
\usepackage{mathrsfs}
\usepackage{amsthm}
\usepackage{rotating}
\usepackage{appendix}
\usepackage[numbers]{natbib}
\usepackage{ifpdf}
\usepackage[T1]{fontenc}
\usepackage{newtxtext}
\usepackage{newtxmath}
\usepackage{textcomp}
\usepackage{xcolor}
\usepackage{lipsum}
\usepackage{pifont}
\usepackage{subcaption}
\usepackage[colorlinks,allcolors=blue]{hyperref}
\usepackage{adjustbox}
\usepackage{booktabs}

\theoremstyle{definition}

\numberwithin{equation}{section}

\title{On the Effectiveness of Neural Operators at Zero-Shot Weather Downscaling}

\author{%
  Saumya Sinha\thanks{Corresponding author: saumya.sinha@nrel.gov}  \\
  National Renewable Energy Laboratory\\
  Golden, Colorado \\
   \\
  \And
  Brandon Benton \\
  National Renewable Energy Laboratory\\
  Golden, Colorado \\
  \AND
  Patrick Emami \\
 National Renewable Energy Laboratory\\
  Golden, Colorado \\
}

\begin{document}








\maketitle

\begin{abstract}
Machine learning (ML) methods have shown great potential for weather downscaling. These data-driven approaches provide a more efficient alternative for producing high-resolution weather datasets and forecasts compared to physics-based numerical simulations. Neural operators, which learn solution operators for a family of partial differential equations (PDEs), have shown great success in scientific ML applications involving physics-driven datasets. Neural operators are grid-resolution-invariant and are often evaluated on higher grid resolutions than they are trained on, i.e., zero-shot super-resolution. Given their promising zero-shot super-resolution performance on dynamical systems emulation, we present a critical investigation of their zero-shot weather downscaling capabilities, which is when models are tasked with producing high-resolution outputs using higher upsampling factors than are seen during training. To this end, we create two realistic downscaling experiments with challenging upsampling factors (e.g., 8x and 15x) across data from different simulations: the European Centre for Medium-Range Weather Forecasts Reanalysis version 5 (ERA5) and the Wind Integration National Dataset Toolkit (WTK). While neural operator-based downscaling models perform better than interpolation and a simple convolutional baseline, we show the surprising performance of an approach that combines a powerful transformer-based model with parameter-free interpolation at zero-shot weather downscaling. We find that this Swin-Transformer-based approach mostly outperforms models with neural operator layers in terms of average error metrics, whereas an Enhanced Super-Resolution Generative Adversarial Network (ESRGAN)-based approach is better than most models in terms of capturing the physics of the ground truth data. We suggest their use in future work as strong baselines.
\end{abstract}

\section{Introduction}
\label{intro}

Downscaling techniques are used to obtain high-resolution (HR) data from their coarse low-resolution (LR) counterparts. The HR data often includes finer details of physical phenomena than the LR data in complex earth systems such as weather. Downscaling provides insights into climate change and its effects, e.g., the small-scale features and detailed information are crucial for analyzing extreme weather events that can only be observed at high resolutions. 
Downscaling can also help upsample medium-range weather forecasts~\cite{jiang2023efficient} and is useful for optimal grid planning and management of renewable resources such as wind energy~\cite{buster2024high, stengel2020adversarial,benton2024super,kurinchi2021wisosuper,ren2023superbench}.

Although earth system processes such as weather and climate can be approximately expressed as systems of partial differential equations (PDEs), solving these models numerically at sufficiently high resolutions for many practical applications is computationally infeasible.
Data-driven downscaling approaches, which promise better efficiency than numerical physics-based solvers, have shown great promise~\cite{ren2023superbench,kurinchi2021wisosuper,yang2023fourier,jiang2023efficient,mikhaylov2024accelerating,buster2024high}. While statistical downscaling methods~\cite{pierce2014statistical,wood2004hydrologic,kaczmarska2014point} have been used traditionally, deep learning techniques, in particular, have gained attention due to their ability to efficiently learn complex relationships from large amounts of data. Moreover, the rapid advancement of deep learning in the computer vision field of super-resolution has been adapted with success for downscaling in the atmospheric sciences~\cite{ren2023superbench,chen2022rainnet,kurinchi2021wisosuper}. 

Neural operators~\cite{kovachki2023neural} have recently been applied to many scientific machine learning (ML) tasks involving the emulation of physical systems. Unlike traditional neural networks, neural operators approximate a mapping between infinite-dimensional function spaces.
For example, neural operators can be used to learn the solution operator for an entire family of PDEs, such as Navier Stokes and Darcy flow~\cite{DBLP:conf/iclr/LiKALBSA21}.
For this application, neural operators are much more efficient than traditional numerical solvers which run on finely discretized grids.
Once trained, neural operators are fast to solve any new instance of the PDE~\cite{kovachki2023neural,DBLP:conf/iclr/LiKALBSA21}. Neural operators have demonstrated the ability to perform \emph{zero-shot super-resolution}~\cite{DBLP:conf/iclr/LiKALBSA21,rahman2023uno,NEURIPS2023_f3c1951b} when emulating physical systems.
That is, they can be trained on coarse resolution data and then tested ``zero-shot'' on a previously unseen fine discretization of a grid. 

Neural operator's ability to perform zero-shot super-resolution raises the question of whether they can be applied to perform \emph{zero-shot weather downscaling}.
Currently, downscaling pipelines train models to map an LR input to an HR output at an upsampling factor (the ratio of the size of the HR grid to the LR grid), and they are evaluated on generating downscaled outputs with the upsampling factor seen during training. 
In zero-shot weather downscaling, a model is trained with a (small) upsampling factor and then the same model is tasked with producing an HR output at an unseen and higher upsampling factor at test time. 
The success of neural operators at zero-shot super-resolution when emulating dynamical systems suggests they hold promise for this task as well.

We design challenging experiments to investigate whether neural-operator-based models have an enhanced ability to perform zero-shot weather downscaling. 
We adapt and expand the learning framework for applying neural operators to this setting as proposed in Yang et al.~\cite{yang2023fourier}.
One of the key difficulties of zero-shot weather downscaling is generalizing to an upsampling factor where the data at the finest spatial scales contains physical phenomena unseen at the highest resolutions seen during training. Fine-scale atmospheric processes such as turbulence and boundary layer dynamics are often unresolved in low-resolution simulations. These simulations also frequently underestimate the intensity of cyclones and storm cells (if they are resolved at all) and smooth over orographic effects due to the coarse representation of topography~\cite{li2021effect,singh2021effects,pryor2012influence}.
We design experiments that aim to test this zero-shot setting by using large upsampling factors (e.g., 8x and 15x) and high target resolutions (e.g., 2km x 2km wind speed data).

Overall, our work investigates the zero-shot downscaling potential of neural operators.
To summarize, our contributions are: 
\begin{enumerate}
    \item We provide a comparative analysis based on two challenging weather downscaling problems, between various neural operator and non-neural-operator methods with large upsampling factors (e.g., 8x and 15x) and fine grid resolutions (e.g., 2km x 2km wind speed). 
    \item We examine whether neural operator layers provide unique advantages when testing downscaling models on upsampling factors higher than those seen during training, i.e., \emph{zero-shot downscaling}. 
    Our results instead show the surprising success of an approach that combines a powerful transformer-based model with a parameter-free interpolation step at zero-shot weather downscaling.
    \item  We find that this Swin-Transformer-based approach mostly outperforms all neural operator models in terms of average error metrics, whereas an Enhanced Super-Resolution Generative Adversarial Network (ESRGAN)-based approach is better than most models in capturing the physics of the system, and suggests their use in future work as strong baselines. However, these approaches still don't capture variations at smaller spatial scales well, including the physical characteristics of turbulence in the HR data. This suggests a potential for improvement in transformer or GAN-based methods and neural-operator-based methods for zero-shot weather downscaling.
\end{enumerate}

\begin{figure}[htbp]
    \centering
    \includegraphics[width=\textwidth]{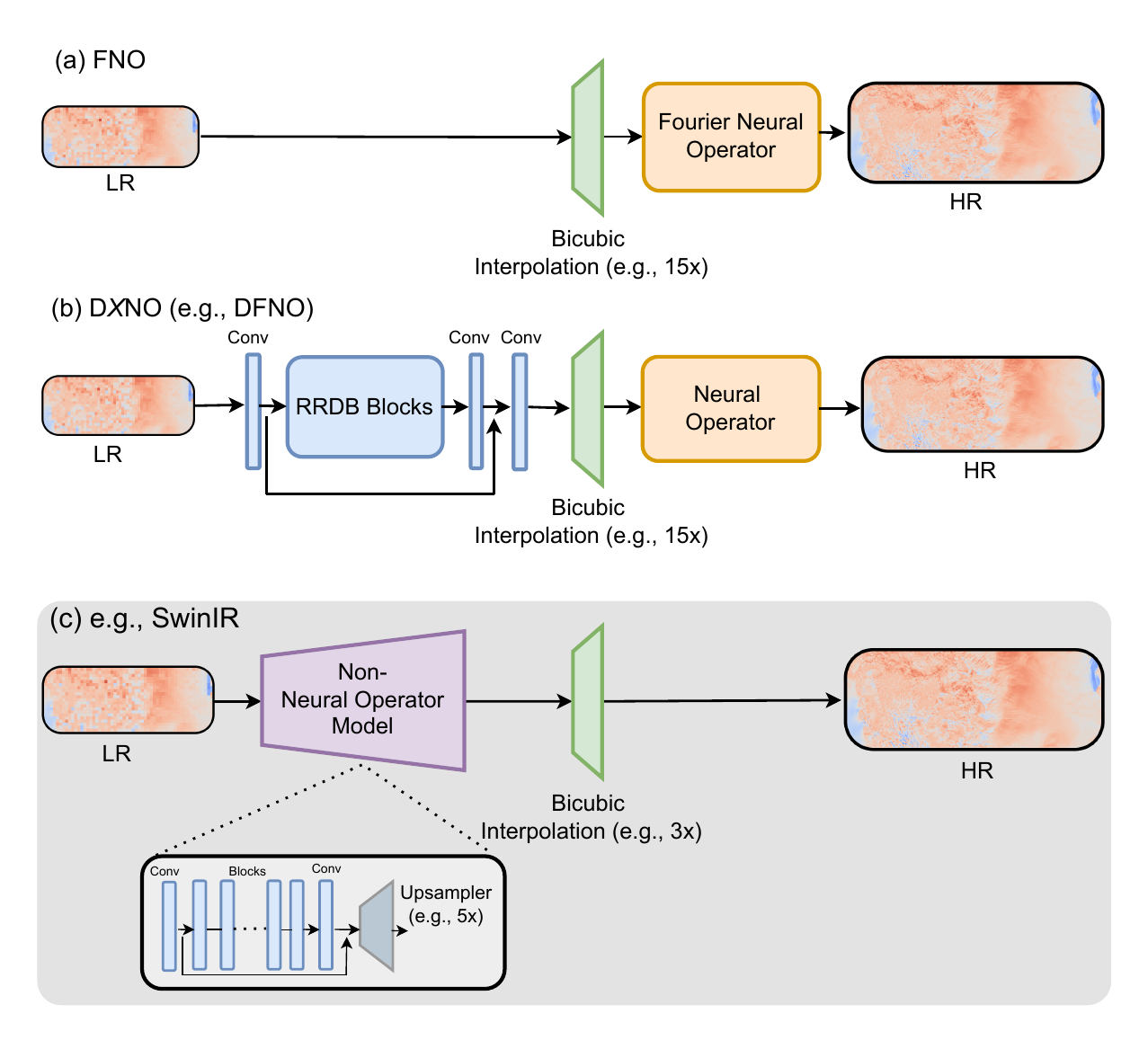}
    \caption{
    Overview of the neural operator and non-neural-operator zero-shot weather downscaling approaches. We show 5x to 15x zero-shot downscaling as an example. (a,b) For neural operators,  the interpolation scale factor is the same as the upsampling factor, e.g., the bicubic interpolation layer upsamples to 5x during training and 15x during evaluation. (c) For regular neural networks (e.g., SwinIR), the model is trained to output at 5x (e.g., using a learnable upsampler such as sub-pixel convolution). At test time, the model generates a 5x output which is then interpolated 3x more to produce the final 15x HR output.
    }
    \label{fig:overall_framework}
\end{figure}
\section{Related Work}
\label{related_work}

\textbf{Weather downscaling with deep learning} 
Deep learning models have recently shown promise at weather downscaling tasks such as precipitation downscaling~\cite{watson2020investigating,chaudhuri2020cligan,harris2022generative}. 
These end-to-end differentiable approaches directly learn to map low-resolution inputs to high-resolution outputs.
The most popular approaches such as the Super-Resolution Convolutional Neural Network (SRCNN)~\cite{dong2015image} are based on architectures introduced by the computer vision community for super-resolution~\cite{wang2020deep}. 
These models are used as key baselines in our experiments, thus, we describe them in more detail in Section~\ref{methodology_experiments}.
Groenke et al.~\cite{groenke2020climalign} have used normalizing flow models~\cite{rezende2015variational} to perform climate downscaling in an unsupervised manner. Harder et al.~\cite{harder2023hard} add physics-based constraints in the learning process of the convolutional neural network (CNN) and generative adversarial network (GAN) models for downscaling variables such as water content and temperature. Our work does not include these constraints as our focus is solely on studying the neural operator layers for the downscaling task. 
Works~\cite{buster2024high, stengel2020adversarial} downscale renewable energy datasets such as wind and solar, as energy system planning depends on high-resolution estimates of these resources. Buster et. al~\cite{buster2024high} use a custom GAN~\cite{stengel2020adversarial} trained on Global Climate Models (GCMs) projections to generate high-resolution spatial and temporal features capturing small-scale details otherwise lost in the coarse GCM models. Benton et al.~\cite{benton2024super} also use the custom GAN~\cite{stengel2020adversarial} for spatiotemporal downscaling of wind data to learn a mapping from low-resolution European Centre
for Medium-Range Weather Forecasts Reanalysis version 5 (ERA5)~\cite{hersbach2020era5} to high-resolution Wind Integration National Dataset
Toolkit (WTK)~\cite{draxl2015wind}. Tran et al.~\cite{tran2020gans} uses Enhanced Super-Resolution Generative Adversarial Network
(ESRGAN)~\cite{wang2018esrgan} for super-resolution of the wind field. Through this work, we also perform downscaling of wind data using ERA5 and WTK datasets, but with the goal of investigating the utility of neural operator models for the weather downscaling task with a specific emphasis on zero-shot downscaling.
\\
\\
\textbf{Related benchmarks for weather downscaling}
SuperBench~\cite{ren2023superbench} introduces super-resolution datasets and benchmarks for scientific applications such as fluid flow, cosmology, and weather downscaling.  Their work compares various deep learning methods and analyzes the physics-preserving properties of these models. 
RainNet~\cite{chen2022rainnet} is one of the first large-scale datasets for the task of spatial precipitation downscaling, spanning over 17 years and covering important meteorological phenomena.  Their work also presents an extensive evaluation of many deep learning models.  
WiSoSuper~\cite{kurinchi2021wisosuper} is a benchmark for wind and solar super-resolution. The dataset released by WiSoSuper is based on the National Renewable Energy Laboratory’s (NREL’s) WTK and National Solar Radiation Database (NSRDB)~\cite{sengupta2018national} datasets. They compare generative models introduced in Stengel et al.~\cite{stengel2020adversarial} with other GAN and CNN-based models.  
In contrast to these benchmarking efforts, our work benchmarks models for weather downscaling but with a focus on neural operator models and zero-shot weather downscaling. 
In a concurrent work focusing on climate downscaling~\cite{prasad2024evaluating}, CNNs, transformers~\cite{alexey2020image}, and neural-operator-based models are compared in terms of their ability to pretrain on diverse climate datasets so as to learn transferrable representations across multiple climate variables and spatial regions.
\\
\\
\textbf{Arbitrary-scale super-resolution} Arbitrary-scale super-resolution (ASSR) has been gaining in popularity~\cite{liu2023arbitrary} in the study of super-resolution in computer vision. ASSR involves training a single deep learning model to super-resolve images to arbitrarily high (potentially non-integral) upsampling factors.
Unlike ASSR, which broadly considers any image datasets relevant to computer vision, our work focuses specifically on zero-shot weather downscaling problems which possess unique challenges due to, for example, multi-scale physical phenomena. 
Super-resolution neural operator (SRNO)~\cite{wei2023super} and the recent Hierarchical Neural Operator Transformer (HiNOTE) model~\cite{luo2024hierarchical} are two advanced deep learning models built with neural operator layers that show promising ASSR performance.
SRNO proposes treating the LR and HR images as approximations of finite dimensional continuous functions with different grid sizes and learns a mapping between them. They introduce an advanced LR image encoding framework and kernel integral layers to learn an expressive mapping. HiNOTE is a hierarchical hybrid neural operator-transformer model. We did not include SRNO and HiNOTE in our experiments as we focus on a simple framework (Figure~\ref{fig:overall_framework}) where downscaling performance can be easily attributed to the neural operator components under study.
\section{Background}
\label{background}
\textbf{Neural operators}
Operator learning models such as neural operators~\cite{kovachki2023neural} are composed of layers that learn mappings between infinite-dimensional function spaces. 
In doing so, they approximately learn a continuous operator, which can be realized at any arbitrary grid discretization of the input and output. Thus, neural operators do not depend on the discretization of the grid they are trained on, and we expect them to generalize to grid resolutions different than the ones they are trained on. Li et al.~\cite{DBLP:conf/iclr/LiKALBSA21} introduced Fourier neural operators (FNOs), expressing neural operators as a combination of linear integral operators to incorporate the non-local properties of the solution operator with Fourier Transform and non-linear local activation functions~\cite{kovachki2023neural}, which helps to model non-linear systems and their high-frequency modes. They show improved performance over convolution-based models for complex non-linear PDEs such as the Navier-Stokes equation. 
With the Fourier layers, the parameters are learned in the Fourier domain, which enables FNOs to be invariant to the grid discretization or resolution. 
Since neural operators such as FNOs learn resolution-invariant approximations of continuous operators, we aim to understand whether this provides advantages for \emph{zero-shot} weather downscaling.

Yang et al.~\cite{yang2023fourier} adapt Fourier neural operators (FNOs)~\cite{DBLP:conf/iclr/LiKALBSA21} to perform downscaling on ERA5 and PDE data. Their proposed model, which they refer to as DFNO, outperforms CNN and GAN-based models. They also evaluate zero-shot downscaling on unseen upsampling factors to observe the model's zero-shot generalization potential. We adapt and expand this downscaling pipeline in our benchmarking study. Our work differs from this paper as we investigate higher upsampling factors (8x and 15x) for training and zero-shot evaluations as opposed to 2x and 4x in their work. We also create a realistic set of experiments that includes LR and HR data sourced from different simulations (ERA5 to WTK downscaling, as described in Section~\ref{era->wtk}) and compare various neural operator approaches against strong baselines including powerful transformers.
\\
\\
\textbf{Weather downscaling}
In weather downscaling, we are given a snapshot of LR weather data (e.g., an image) with a goal of upsampling this data to a higher target resolution.
Mathematically, in the standard downscaling problem we have the LR input grid $\mathbf{x} \in \mathbb{R}^{h \times w \times c}$, and a target, HR output, $\mathbf{y} \in \mathbb{R}^{h' \times w' \times c}$, where $h,w \in \mathbb{N}$, $c$ is the number of atmospheric variables, and $h \times w$ is a lower resolution than $h' \times w'$. Deep-learning-based downscaling techniques introduced in Section~\ref{related_work} learn an approximation $f$ between two finite-dimensional vector spaces $f: \mathbb{R}^{h \times w \times c} \rightarrow \mathbb{R}^{h' \times w' \times c}$.
We refer to the setting where models are trained and tested on the same upsampling factor as \emph{standard} weather downscaling.
In this work, we restrict our focus to only static downscaling problems, i.e., each snapshot represents a single instant in time.
\\
\\
\textbf{Zero-shot weather downscaling} 
In our work, we wish to evaluate the extent to which downscaling models built with resolution-invariant neural operator layers generalize when tested on previously unseen, higher upsampling factors compared to approaches without such layers. The simplest way to obtain an HR image at any arbitrarily fine discretization is a non-learned interpolation scheme such as bicubic interpolation.

We are looking into neural-operator-based downscaling models that learn a mapping $\mathcal{G}^\dagger: \mathbb{R}^{h \times w \times c} \rightarrow \mathcal{U}$ from $\mathbf{x} \in \mathbb{R}^{h \times w \times c}$ to a function $u \in \mathcal{U}$. We aim to obtain HR outputs $\mathbf{y} \in \mathbb{R}^{h' \times w' \times c}$  from a discretization of $u$, where $\mathcal{U}$ is an infinite-dimensional function space~\cite{DBLP:conf/iclr/LiKALBSA21,yang2023fourier}. 
A neural-operator-based downscaling framework~\cite{yang2023fourier}
(Figure~\ref{fig:overall_framework}b) learns a parametric approximation of a mapping from the finite-dimensional LR input space to the infinite-dimensional space, $G_{\theta}(\mathbf{x}): \mathbb{R}^{h \times w \times c} \rightarrow \mathcal{U}$, as an approximation of $\mathcal{G}^\dagger$ such that $G_{\theta}(\mathbf{x}) := \mathcal{F}_{\theta}(T^{-1}(f_{\theta}(\mathbf{x})))$, with ${\theta}$ as the parameters of the model. It is comprised of (a) neural network layers that first learn to map LR inputs to an embedding vector, $f_{\theta}: \mathbb{R}^{h \times w \times c} \rightarrow \mathbb{R}^d$, (b) a discretization inversion operator that converts the vector to a function ($e \in \mathcal{E}$) with $T^{-1}: \mathbb{R}^d \rightarrow \mathcal{E}(D; \mathbb{R}^{d_e})$, and (c) neural operator layers $\mathcal{F}_{\theta} : \mathcal{E} \rightarrow \mathcal{U}$ that learn to map the function to another function, which can be discretized to produce the HR output $\mathbf{y} \in \mathbb{R}^{h' \times w' \times c}$. We refer to these approaches as Downscaling NO models (e.g. DFNO~\cite{yang2023fourier}). In order to use vanilla FNOs without resolution-dependent neural network layers (a)  (as seen in Figure~\ref{fig:overall_framework}(a)), we learn $G_{\theta}(\mathbf{x}) := \mathcal{F}_{\theta}(T^{-1}(\mathbf{x}))$. 
Several improvements have since been proposed over FNOs ~\cite{guibas2021adaptive,rahman2023uno,NEURIPS2023_f3c1951b} which we include in our downscaling study and describe in further detail later (Section~\ref{methodology_experiments}).
\\

\section{Methodology}
\label{methodology_experiments}
We use two experimental setups to compare the performance of neural operators and non-neural-operator-based methods at both standard and zero-shot weather downscaling. First, we downscale ERA5 data, where we learn a mapping from coarsened LR ERA5 to HR ERA5. In our second experiment, we downscale from LR ERA5 to HR WTK. We expect the second task to be more challenging as it presents a more realistic downscaling scenario where the LR inputs belong to a different simulation than the HR data. Thus, we do not assume the LR is a coarsened version of the HR~\cite{ren2023superbench,benton2024super}. The coarse simulation in this case differs from the fine simulation in terms of modeling assumptions and physics. The performance of our downscaling models for this complex task provides important insights into their capability to infer unseen fine-scale physical phenomena and to be used in an operational context. 
\\
\\
\textbf{Downscaling Neural Operator models}
We compare (vanilla) FNO with Downscaling FNO (DFNO), Downscaling U-shaped Neural Operator~\cite{rahman2023uno} (DUNO), Downscaling Convolutional Neural Operator~\cite{NEURIPS2023_f3c1951b} (DCNO), and Downscaling Adaptive Fourier Neural Operator~\cite{guibas2021adaptive} (DAFNO).
The Downscaling (D) models are based on Yang et al.~\cite{yang2023fourier} (as described in Section~\ref{background}) with FNO, UNO, CNO, and AFNO as the neural operator layers in the modeling framework. 
We show details of this model in Figure~\ref{fig:overall_framework}(b).
The low-resolution (LR) image first passes through a set of Residual-in-Residual Dense Block (RRDB) blocks, where an RRDB block is composed of multiple levels of residual and dense networks as introduced in ESRGAN~\cite{wang2018esrgan}.
 Then, the embedding is interpolated corresponding to the upsampling factor using bicubic interpolation to obtain a high-resolution output. Finally, this goes through neural operator layers to produce the final downscaled HR image. We can think of this last stage as a post-processing step over the features extracted by the RRDB layers followed by the interpolation. The number of RRDB blocks is a hyperparameter tuned during training. All the Downscaling (D) models are trained with the mean-squared-error (MSE) loss. 
To perform zero-shot downscaling with either the FNO or D\textit{X}NO (e.g. DFNO) models at the test time on higher upsampling factors than the ones seen during the training, we use the interpolation layer to increase the resolution by the corresponding higher upsampling factor. 

The four neural operator models we compare are:
\begin{enumerate}
\item \textbf{FNO}~\cite{DBLP:conf/iclr/LiKALBSA21} FNO uses a combination of
linear integral operators and non-linear local activation functions to learn the operator mapping on complex PDEs such as the Navier-Stokes equation. We follow the original implementation of FNO from~\cite{DBLP:conf/iclr/LiKALBSA21} for the FNO model. Figure~\ref{fig:overall_framework}(a) shows the vanilla FNO model as used in our downscaling pipeline where FNO is post-processing the interpolation output.
\item \textbf{UNO}~\cite{rahman2023uno} is introduced as a deep and memory-efficient architecture that allows for faster training than FNOs. 
This neural operator has a U-shaped architecture, comprising an encoder-decoder framework built with neural operator layers to learn mappings between function spaces over different domains. 
\item \textbf{AFNO}~\cite{guibas2021adaptive}, a transformer model that uses FNOs for efficient token-mixing instead of the traditional self-attention layers. AFNOs have been used as an integral part of FourCastNet~\cite{pathak2022fourcastnet}, a data-driven weather forecasting model. 
\item \textbf{CNO}~\cite{NEURIPS2023_f3c1951b} adapts CNNs for learning operators and processing the inputs and outputs as function spaces. They adopt a UNet~\cite{ronneberger2015u}-based modeling framework to learn a mapping between bandlimited functions~\cite{vetterli2014foundations}, which supports resolution invariance within the captured frequency band and can help them to learn operators that reduce aliasing errors~\cite{bartolucci2024representation} which occurs when the neural operator models try to learn a continuous operator on a finite, discretized grid.
\end{enumerate}

The hyperparameters and model training details are presented in Appendix~\ref{hyperparameter_details}.
Additionally, we also conducted a detailed analysis of how the frequency cutoff hyperparameter affects the downscaling performance of the FNO, see Appendix~\ref{fno_mode_analysis}.
 
\textbf{Baseline models} 
We compare all the neural operator models with five baselines: (1) bicubic interpolation, (2) SRCNN, (3) ESRGAN, (4) EDSR, and (5) SwinIR.
Super-Resolution Convolutional Neural Network or SRCNN~\cite{dong2015image} is the first CNN-based model to perform single image or spatial super-resolution. SRCNN first upsamples the LR input with bicubic interpolation followed by lightweight CNN layers to obtain the HR image. 
Enhanced Deep Super-Resolution Network (EDSR)~\cite{lim2017enhanced} introduces deep residual CNN networks to do super-resolution where the CNN layers are followed by an upsampling block performing sub-pixel convolution~\cite{shi2016real}.
Enhanced Super-Resolution Generative Adversarial Network (ESRGAN)~\cite{wang2018esrgan} is a GAN-based architecture where the generator is composed of many Residual-in-Residual Dense Block (RRDB) blocks. The model is trained with pixel and perceptual loss along with adversarial loss, the perceptual loss minimizes the errors in the feature space and helps improve the visual quality of the generated super-resolved image~\cite{ledig2017photo}.
Swin Transformer for Image Restoration (SwinIR)~\cite{liang2021swinir} has the advantages of both the CNN and Swin transformer~\cite{liu2021swin} layers. It captures long-range dependencies and learns robust features to improve super-resolution performance with the residual Swin Transformer blocks (RSTB) which is composed of many Swin Transformer layers stacked together with residual connections. 

 We refer to the implementation of these models as released in the SuperBench~\cite{ren2023superbench} work, hyperparameter details are added in Appendix~\ref{hyperparameter_details}. Unlike neural operators, these models have architectures that expect their inputs and outputs to have the same grid resolution at both training and test time. Thus, we add a bicubic interpolation module on the output obtained from these models to produce outputs at higher upsampling factors than seen during training for our zero-shot downscaling experiments (Figure~\ref{fig:overall_framework}(c)). 
 
 See Appendix~\ref{parameter_details} comparing the parameter count and training wall-clock time of all the neural operator and non-neural-operator-based models. Notably, we tested the vanilla FNO downscaling framework (Figure~\ref{fig:overall_framework}(a)), by moving the bicubic interpolation module after the FNO layers (as in baseline models, Figure~\ref{fig:overall_framework}(c)), which led to worse performance compared to our proposed pipeline with interpolation before the FNO layers. 
 
\subsection{ERA5 to ERA5 downscaling}
For our first experiment, we downscale coarsened LR ERA5 to HR ERA5. 
We use the ERA5 dataset for the entire globe at 25-km spatial resolution.
We compare all models using two downscaling paradigms:
\begin{enumerate}
\item Standard downscaling: We train and test all the neural operators and baseline models with \emph{the same} upsampling factor of 8x.  An upsampling factor of 8x maps LR images of size 90x180 to HR outputs of size 720x1440. 
\item Zero-shot downscaling: We first train all the models with an upsampling factor of 4x. Then, during testing, we observe their ability to produce outputs at a \emph{higher} 8x upsampling factor. 
\end{enumerate}

\textbf{Dataset details} 
We use the European Centre for Medium-Range Weather Forecasts Reanalysis version (ERA5) dataset released as a part of the SuperBench~\cite{ren2023superbench} paper. The data is at a 0.25-degree (25km) grid resolution over the globe, i.e., each image has size 720x1440. We have three atmospheric variables (three channels): (1) wind speed, $\sqrt{u^2 + v^2}$, $u$ and $v$ being the two components of wind velocity at 10m from the surface, (2) temperature at 2m from the surface, and (3) total column water vapor. This ERA5 dataset consists of image snapshots at 24-hour intervals over an eight year period. Years 2008, 2010, 2011, and 2013 are used for training while 2012 and 2007 are reserved as a validation set for tuning hyperparameters. The years 2014 and 2015 are set aside for testing. Following Ren et al.~\cite{ren2023superbench}, we extract eight patches of size 64x64 (for the zero-shot downscaling) from each image to obtain HR images for training. The LR images are created by coarsening the HR patches with bicubic interpolation. We normalize each channel separately with a mean and standard deviation before training. See Appendix~\ref{data_details} for further details on the training process.


\subsection{ERA5 to WTK downscaling}
\label{era->wtk}
For this second experiment, we focus on downscaling from LR ERA5 to HR WTK. It should be noted that the LR data in this setup is not obtained by coarsening the HR data but comes from another simulation. We include this experiment to observe the performance of neural operators in a challenging and more realistic setup; for e.g., where ERA5 serves as boundary conditions for dynamical downscaling with a Numerical Weather Prediction model~\cite{benton2024super}. The HR WTK dataset is available over two regions in the US at a 2-km resolution~\cite{benton2024super} and we use two variables: the $u$ and $v$ components of the wind velocity at 10m from the surface for this task. 
We perform the following experiments with the ERA5 to WTK downscaling setup:
\begin{enumerate}
\item Standard downscaling: We train and test all the neural operators and the baseline models with an upsampling factor of 5x, to go from 30-km to 6-km. The LR and the HR sizes are 53x53 and 265x265 for one region and 40x106 and 200x530 for the other region.
\item Zero-shot downscaling: For the zero-shot setup, we use the models trained with the 5x upsampling and evaluate them on an upsampling factor of 15, going from 30-km to 2-km. While the LR sizes are the same as above, the HR sizes for the zero-shot case are 795x795 for one region and 600x1590 for the second region.
\end{enumerate}

\begin{figure}[htbp]
    \centering
    \includegraphics[width=0.7\textwidth]{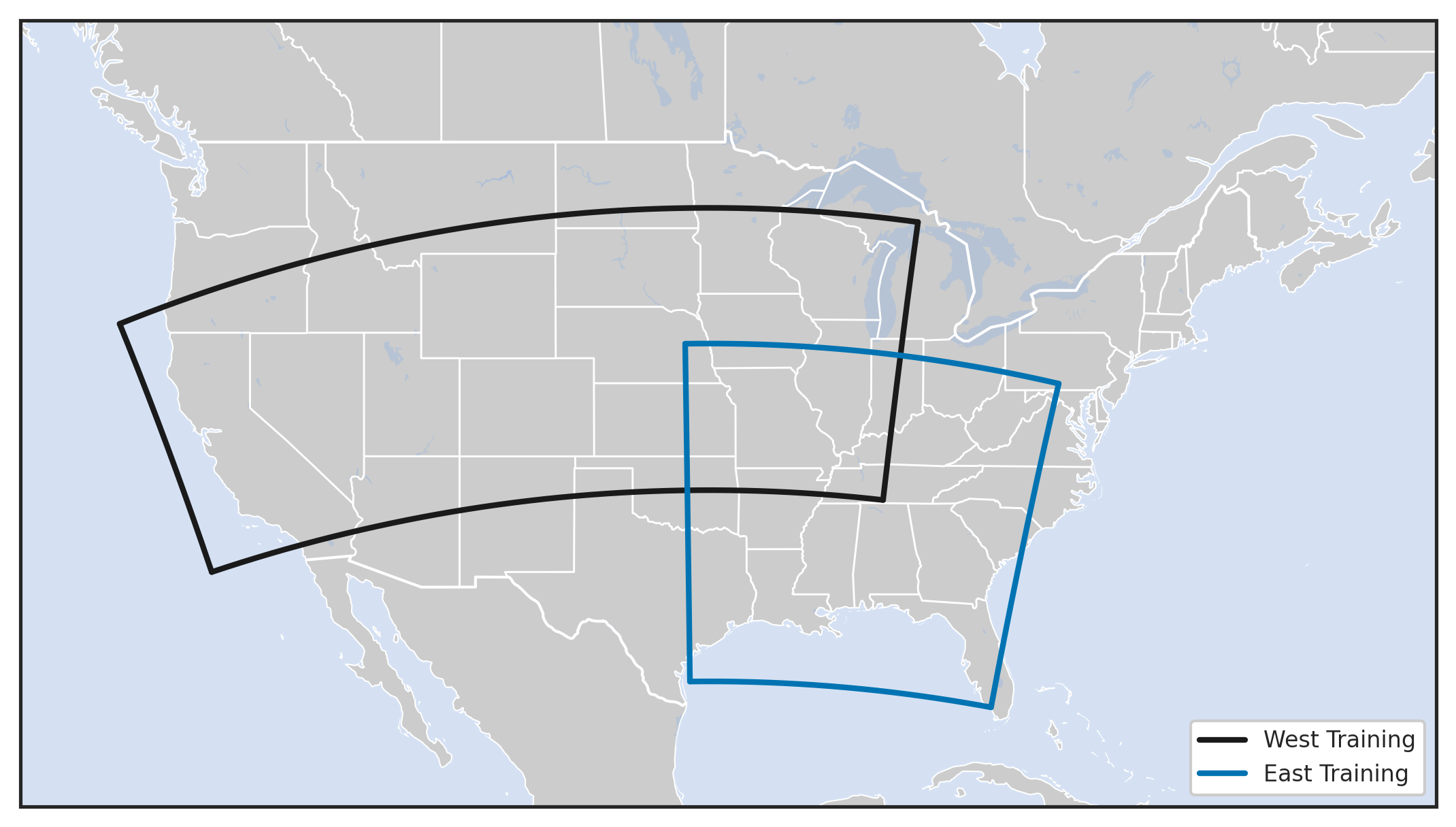}
    \caption{Figure shows the two regions used for the ERA5 to WTK downscaling experiments. The 600x1600 (800x800) region is shown in black (blue). We use NREL's rex~\cite{rossol2021rex} tools to rasterize the WTK dataset.}
    \label{fig:wtk_regions}
\end{figure}

\textbf{Dataset details} 
We use the National Renewable Energy Laboratory's (NREL's) Wind Integration National Dataset Toolkit~\cite{draxl2015wind} (WTK) as the ground truth dataset. WTK has a spatial resolution of 2-km and a temporal resolution of 1-hour. We get LR images from the ERA5 dataset~\cite{hersbach2020era5}, available at 30-km ($\sim$0.28 degree) spatial resolution and 1-hour temporal resolution.  This data has two channels, the $u$ and $v$ components of the wind velocity at 10m from the surface. We have paired ERA5 and WTK datasets over two regions in the US, with image sizes $\sim$800x800 and $\sim$600x1600 respectively (see Figure~\ref{fig:wtk_regions} and~\cite{benton2024super} for more details). 

Models are trained to map from coarse 30-km ERA5 to fine 6-km WTK with a 5x upsampling factor. This HR data is created by coarsening the WTK grid from 2-km to 6-km resolution. We realign, i.e. regrid 30-km ERA5 to the 6-km WTK coarsened grid using inverse distance weighted interpolation. For zero-shot experiments, we map the 30-km low-resolution ERA5 to the original 2-km resolution WTK (a 15x upsampling) for evaluation. The year 2007 is split 80/20 between the training and validation. We keep the year 2010 for testing. A single model is trained for both regions. While training, we ensure that every batch has an equal number of LR and HR pairs from both regions.  We extract patches of size 160x160 from the WTK image tiles to obtain the HR images and the corresponding coarse patches (32x32) from ERA5 image tiles as the LR images for training~\cite{benton2024super}. This patch size is a hyperparameter tuned on the validation set. We normalize each channel separately with a mean and standard deviation before training. See Appendix~\ref{data_details} for further details on the training process.


\subsection{Evaluation metrics}
Our study quantitatively analyzes model performance using the following:
\begin{enumerate}
\item \textbf{Error metrics}: We use four pixel-level error measures: mean-squared-error (MSE), mean-absolute-error (MAE), $L_\infty$ norm (IN), and the Peak signal-to-noise ratio (PSNR~\cite{ren2023superbench}). IN is the maximum pixel error between two images and it informs us about the tails of the pixel error distribution~\cite{ren2023superbench}.
\item \textbf{Energy spectrum}: We plot the kinetic energy spectrum~\cite{kolmogorov1991dissipation} for each model which shows a distribution of energy across various wavenumbers~\cite{buster2024high, stengel2020adversarial,kurinchi2021wisosuper,benton2024super}. These are normalized kinetic energy plots, with wavenumbers measured relative to the domain (or the spatial region) size. We compare each of the models with the energy curve for the ground truth HR. These plots describe how well the models capture physically realistic variations at smaller spatial scales in their downscaled outputs, for example, providing information about the physical characteristics of the turbulence of wind flow captured in the model outputs.
\end{enumerate}
\section{Evaluation}
\label{evaluation}
\subsection{ERA5 to ERA5 downscaling}
\label{era_ds}
\textbf{Error Metrics} We show the results for the ERA5 to ERA5 standard downscaling experiments and zero-shot experiments in Table~\ref{table:era5_8x}. Standard downscaling compares models trained and evaluated with an upsampling factor of 8x. 
We observe that SwinIR outperforms every other model in terms of MSE, MAE, IN, and PSNR. DCNO is a close second and the best-performing neural operator model. The DFNO model shows improved results over the vanilla FNO indicating the advantage of adding convolutional RRDB layers that learn spatial domain features useful for downscaling (as shown in ESRGAN~\cite{wang2018esrgan}). Table~\ref{table:era5_8x} zero-shot results show a performance comparison between models trained on a 4x upsampling factor but evaluated zero-shot on generating 8x upsampled HR outputs. The zero-shot experiments show that the SwinIR is still the best-performing model. While DUNO is best among the neural operator models at zero-shot ERA5 to ERA5 downscaling, DCNO performs much worse than it did on standard downscaling. All the neural operator models are better than bicubic and SRCNN for both the standard and zero-shot downscaling.
We also include downscaling results on the temperature and total column water vapor variables for the ERA5 to ERA5 experiments in Appendix~\ref{appendix_era} (Tables~\ref{table:era5_temp_8x},~\ref{table:era_8x_tchw}) and observe similar relative model performance, with SwinIR outperforming every other model in both the standard and zero-shot downscaling in terms of all the metrics. 
\begin{table}[t]
    \centering
    \caption{\textbf{ERA5 to ERA5 wind speed downscaling results}. MSE has units $(m/s)^2$, MAE $m/s$ and IN $m/s$.  We bold the best-performing model among all the models and underline the best-performing neural operator model. Results for the other channels are added to the Appendix~\ref{appendix_era}. 
    \label{table:era5_8x} }
    \begin{adjustbox}{max width=\columnwidth}
    \begin{tabular}{lccccccccc}
        \toprule
        & & \multicolumn{4}{c}{\textbf{Standard Downscaling}}           & \multicolumn{4}{c}{\textbf{Zero-shot Downscaling}}           \\ \cmidrule(l){2-10}
        & is NO? & MSE $\downarrow$ & MAE $\downarrow$ & IN $\downarrow$ & PSNR $\uparrow$ & MSE $\downarrow$ & MAE $\downarrow$ & IN $\downarrow$ & PSNR $\uparrow$ \\
        \midrule
        bicubic & \ding{55}  & 1.23  & 0.73 & 14.82  & 27.53   & 1.23  & 0.73 & 14.82  & 27.53     \\
        SRCNN & \ding{55} & 1.14  & 0.7  & 14.75  & 27.83   & 1.06  & 0.67  & 14.56  & 28.18   \\
        ESRGAN & \ding{55} & 1.29  & 0.75  & 15.43  &  27.3  & 0.85  & 0.6  & 14.51  & 29.1   \\
        EDSR & \ding{55} & 0.51  & 0.44 & 13.6 & 31.33  & 0.54  & 0.45 & 13.66 & 31.1     \\
        SwinIR & \ding{55} & \textbf{0.37} & \textbf{0.38}  & \textbf{12.12}  & \textbf{32.79}  & \textbf{0.51} & \textbf{0.44}  & \textbf{13.22}  & \textbf{31.33}          \\
        \midrule
        FNO & \ding{51} & 0.91  & 0.66  & 14.53  & 28.84  & 0.71  & 0.54  & 14.04  & 29.9  \\
        DFNO & \ding{51} & 0.7  & 0.54  & 12.8  & 29.94  & 0.63  & 0.5  & 13.33  & 30.43  \\
        DUNO & \ding{51} & 0.69  & 0.53 & 13.36 & 30.04  & \underline{0.63}  & \underline{0.5} & \underline{13.56} & \underline{30.44}   \\
        DAFNO & \ding{51} & 0.65  & 0.51  & 13.77  & 30.29 & 0.66  & 0.52  & 14.01  & 30.23     \\
        DCNO & \ding{51} & \underline{0.45}  & \underline{0.43}  & \underline{12.98} & \underline{31.93}  & 0.92  & 0.65  & 14.89 & 28.76    \\
        \bottomrule
    \end{tabular}
    \end{adjustbox}
\end{table}

\begin{figure}[htbp] 
    \centering
    \begin{subfigure}{0.75\textwidth} 
        \centering
        \includegraphics[width=\textwidth]{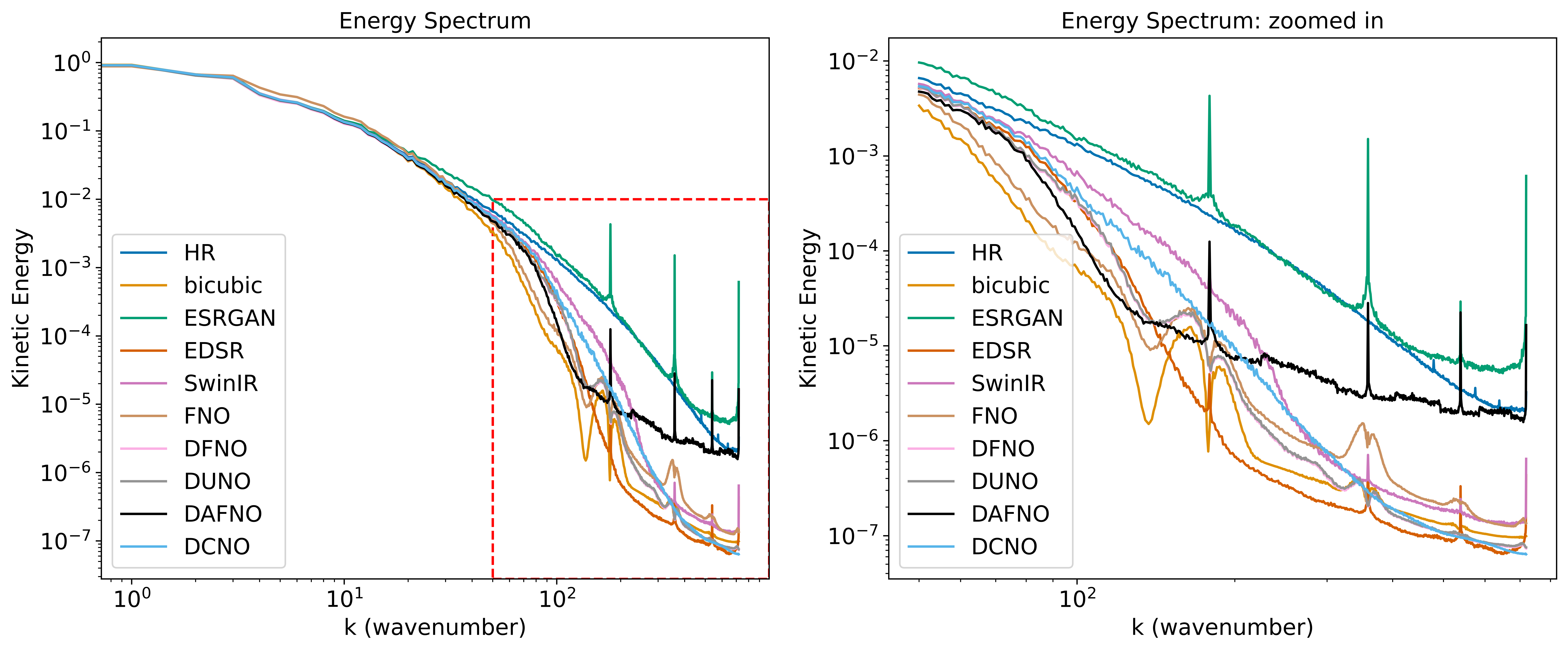}  
        \caption{}
         \label{fig:ke_era_spectrum}
    \end{subfigure}
    \hfill
    \begin{subfigure}{0.75\textwidth} 
        \centering
        \includegraphics[width=\textwidth]{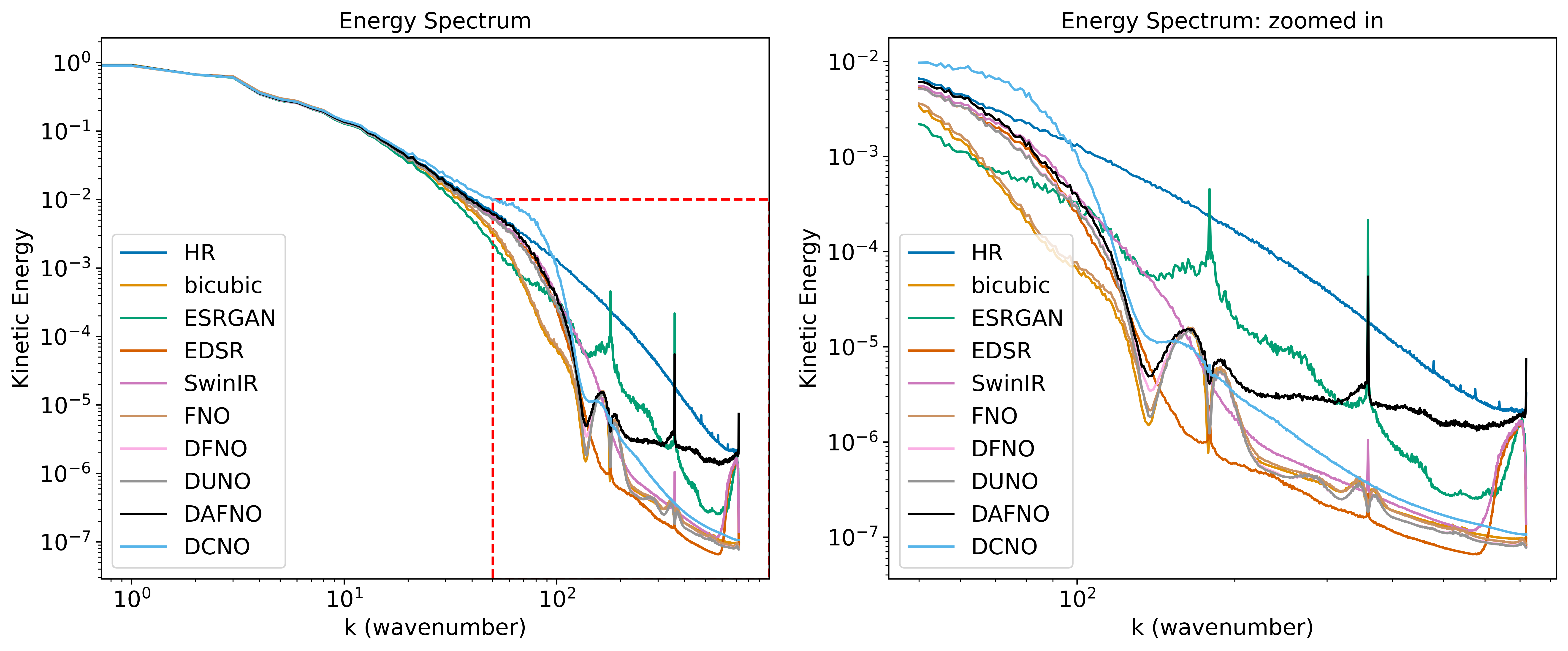} 
        \caption{}
    \label{fig:ke_era_zeroshot_spectrum}
    \end{subfigure}
    \caption{Figures (a) and (b) show kinetic energy spectrum plots for ERA5 standard downscaling and zero-shot downscaling respectively. \textit{Kinetic Energy is normalized and wavenumber is measured relative to the domain size.} We add a zoomed-in plot (right) beside the main plot to zoom in on the key region of interest. We observe that ESRGAN appears to
be the best in capturing the ground truth energy spectra across all wavenumbers for standard downscaling (a) and at
medium-to-high wavenumbers for zero-shot downscaling (b). DAFNO is the best performing at the highest wavenumbers in (b). 
    }
    \label{fig:ke_era_plots}
\end{figure}
\begin{figure}[htbp]
    \centering
    \includegraphics[width=1\textwidth]{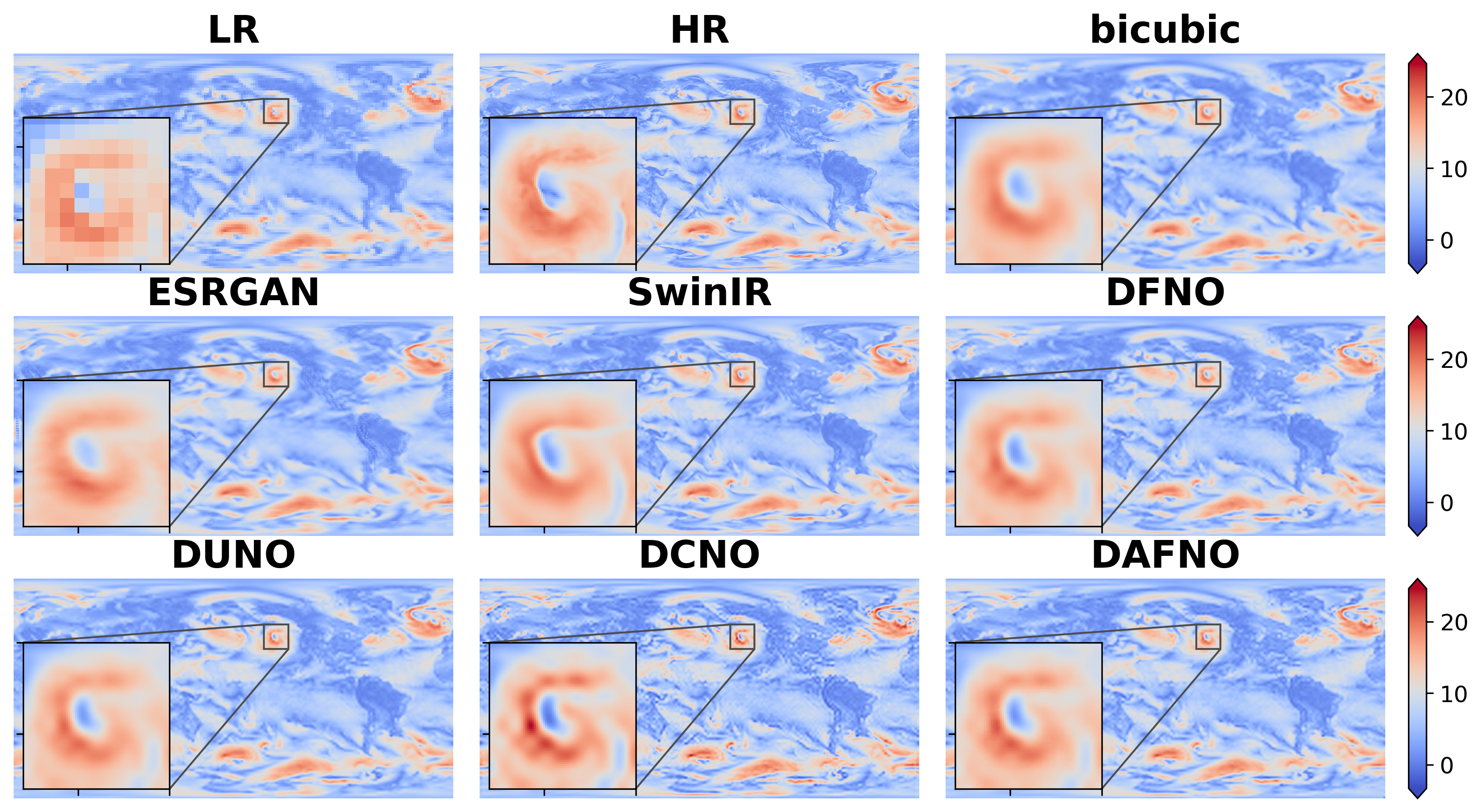}
    \caption{ERA5 wind speed visualizations in $m/s$ generated from the zero-shot downscaling. We zoom in on a small region for better comparison. SwinIR captures better and finer details of the HR image, over neural operator models, especially in the zoomed-in region. It is also better over regions with complex terrain (e.g. the mountain ranges in North and South America).}
    \label{fig:viz_era_zeroshot}
\end{figure}

\textbf{Takeaways:} SwinIR outperforms every other model in both the standard and zero-shot downscaling in terms of all the metrics. While we expected neural operator models to be better, especially in the zero-shot experiments, they don't perform as well as SwinIR.
DUNO achieves the best zero-shot downscaling performance amongst the neural operator models, while DCNO achieves much lower errors than DUNO in the standard downscaling setting.
\\
\\
\textbf{Energy Spectrum} Figure~\ref{fig:viz_era_zeroshot} shows zero-shot downscaled wind speed for the SwinIR, ESRGAN, DFNO, DUNO, DCNO, and DAFNO models, alongside the LR, HR, and bicubic interpolated images. We observe that SwinIR appears to learn fine-scale features closest to the HR or the ground truth image. We refer to the energy spectrum plots in Figures~\ref{fig:ke_era_spectrum} and~\ref{fig:ke_era_zeroshot_spectrum} to show the kinetic energy distributions as functions of wavenumber, 
across all the downscaling models. 
For the standard downscaling case in Figure~\ref{fig:ke_era_spectrum}, ESRGAN matches
the HR spectrum at all wavenumbers,
 DAFNO is second to ESRGAN at high wavenumbers, even though DAFNO is behind the DCNO model in terms of the error metrics. 
Figure~\ref{fig:ke_era_zeroshot_spectrum} demonstrates the energy spectrum for the zero-shot downscaling scenario. SwinIR best captures the physical properties of the ground truth at low-to-medium wavenumbers, ESRGAN is better at medium-to-high wavenumbers but DAFNO is the best at the highest wavenumbers, even though they underestimate the energy content at higher wavenumbers. 
DCNO matches the HR curve for lower wavenumbers but falls behind ESRGAN at higher wavenumbers. 
We observe that SwinIR, ESRGAN, and EDSR produce peaks at the very high-end wavenumbers but the neural operator models except DAFNO do not introduce this high-end noise. Moreover, DUNO and DFNO seem to have similar intermediate peaks as bicubic interpolation, but that's not the case with DAFNO and DCNO. We suggest this points to the difference in spatial domain features learned by DAFNO and DCNO which leads to a change in their spectrum. We also suspect the intermediate artifacts or spikes in ESRGAN and DAFNO (in both Figures~\ref{fig:ke_era_spectrum} and~\ref{fig:ke_era_zeroshot_spectrum}) might be caused by aliasing.

\textbf{Takeaways:} All the models underestimate the energy content at the higher wavenumbers, for the case of zero-shot downscaling. 
For the highest wavenumbers, or the dissipation range, this underestimation is most significant (for all the models except DAFNO). The inherent length dependence of the dissipation range makes zero-short downscaling a challenge for even the best-performing models.  
This is not surprising, as, for example, it is challenging for the models to fill in smaller-scale physical features if they do not see this level of detail when training on smaller upsampling factors. ESRGAN appears to be the best in capturing the ground truth energy spectra across all wavenumbers for standard downscaling and at medium-to-high wavenumbers for zero-shot downscaling. However, DAFNO outperforms all the models at the highest wavenumbers for the zero-shot downscaling case.



\subsection{ERA5 to WTK downscaling}
\label{era_wtk}
\textbf{Error Metrics} We show the results for the ERA5 to WTK downscaling in Table~\ref{table:wtk_5x}. As discussed in Section~\ref{era->wtk}, using ERA5 as the LR for this setup makes it more challenging as the LR data is obtained from a different simulation than the HR data rather than using a coarsened version of the HR data.
Table~\ref{table:wtk_5x} shows the (1) standard downscaling results obtained from evaluating the models mapping ERA5 to WTK for a 5x upsampling factor and (2) zero-shot downscaling results where we evaluate the models trained on 5x upsampling to generate HR outputs at a 15x upsampling factor. SwinIR remains the best-performing model in both setups. DCNO achieves the best standard downscaling scores among the neural operators but is poor at zero-shot downscaling, as can be seen in Table~\ref{table:wtk_5x} zero-shot results, where it performs worse than bicubic interpolation. We also observe DAFNO to be performing worse than bicubic interpolation at zero-shot downscaling. EDSR is a close second to SwinIR in both experiments. DUNO performs better than the other neural operators (and bicubic as well as SRCNN) at zero-shot downscaling.
\begin{table}[t]
    \centering
    \caption{\textbf{ERA5 to WTK wind speed downscaling results}. MSE has units $(m/s)^2$, MAE $m/s$ and IN $m/s$. We aggregate the error metrics over u and v wind velocity channels. We bold the best-performing model among all the models and underline the best-performing neural operator model.}
    \label{table:wtk_5x}
    \begin{adjustbox}{max width=\columnwidth}
    \begin{tabular}{lccccccccc}
        \toprule
          & & \multicolumn{4}{c}{\textbf{Standard Downscaling}}           & \multicolumn{4}{c}{\textbf{Zero-shot Downscaling}}           \\ \cmidrule(l){2-10}
        & is NO? & MSE $\downarrow$ & MAE $\downarrow$ & IN $\downarrow$ & PSNR $\uparrow$ & MSE $\downarrow$ & MAE $\downarrow$ & IN $\downarrow$ & PSNR $\uparrow$ \\
        \midrule
        bicubic & \ding{55}  & 3.56  & 1.18 & 12.87  & 18.4   & 4.07  & 1.25 & 16.59  & 19.91  \\
        SRCNN & \ding{55} & 3.16  & 1.11  & 12.62  & 18.83 & 3.65  & 1.18  & 16.39  & 20.31    \\
        ESRGAN & \ding{55} & 2.75  & 1.05  &  13.06  & 19.27   & 3.12  & 1.11  & 15.96  &  20.8  \\
        EDSR & \ding{55} & 2.46  & 0.98 & 11.89 & 19.85   & 2.92  & 1.05 & 15.61 & 21.2     \\
        SwinIR & \ding{55} & \textbf{2.29} & \textbf{0.94}  & \textbf{11.71}  & \textbf{20.12}  & \textbf{2.73} & \textbf{1.02}  & \textbf{15.34}  & \textbf{21.43}      \\
        \midrule
        FNO & \ding{51} & 5.69  & 1.94  & 14.45 & 14.48 & 5.45  & 1.89  & 17.82  & 16.76    \\
        DFNO & \ding{51} & 3.04  & 1.26  & 12.56  & 17.86   & 3.53  & 1.33  & 16.14  & 19.4  \\
        DUNO & \ding{51} & 2.81  & 1.09 & 12.11 & 18.97  &  \underline{3.3}  &  \underline{1.16} &  \underline{15.85} &  \underline{20.43}  \\
        DAFNO & \ding{51} & 2.71  & 1.02  & 12.12  & 19.47  & 4.17  & 1.19  & 17.5  & 19.51   \\
        DCNO & \ding{51} &  \underline{2.47}  &  \underline{0.99}  &  \underline{11.77} &  \underline{19.79}  & 4.66  & 1.32  & 17.32 & 19.51   \\
        \bottomrule
    \end{tabular}
    \end{adjustbox}
\end{table}

\begin{figure}[htbp] 
    \centering
    \begin{subfigure}{0.75\textwidth} 
        \centering
        \includegraphics[width=\textwidth]{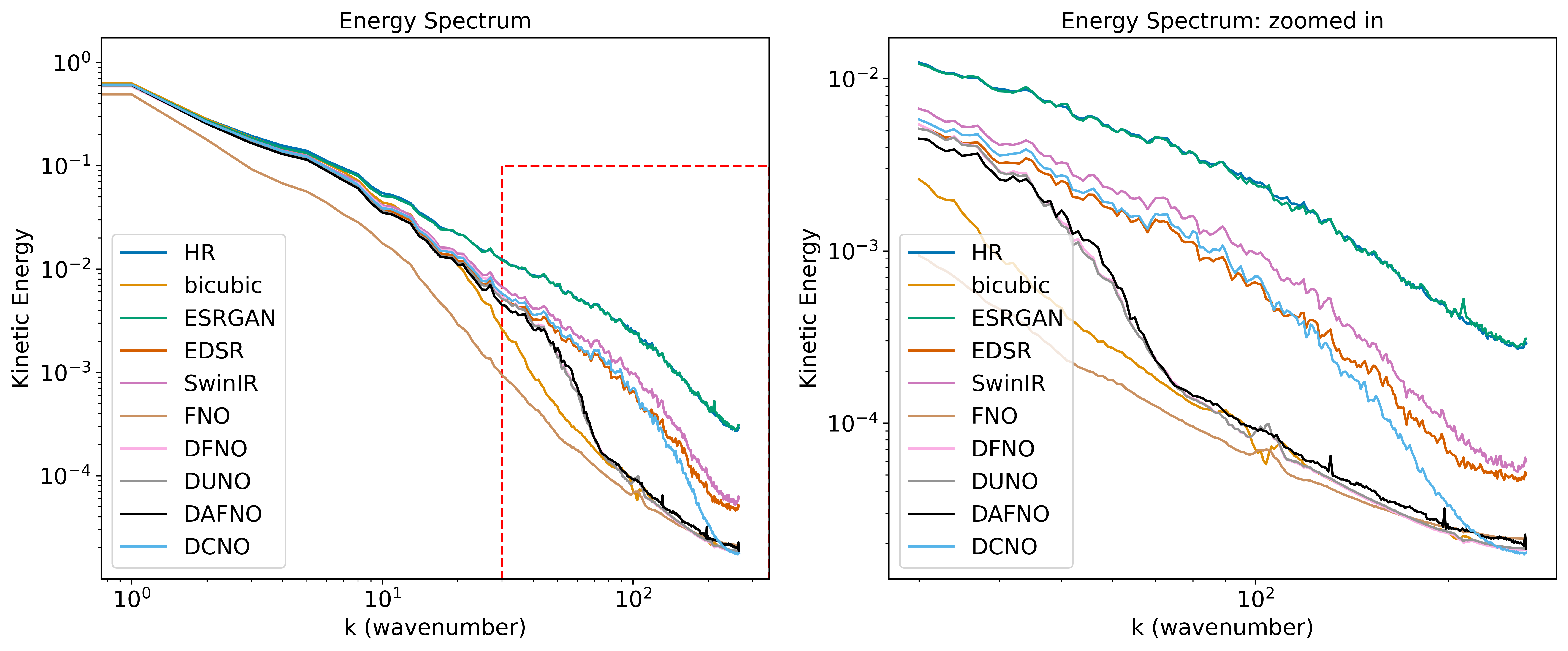}  
        \caption{}
         \label{fig:ke_wtk_spectrum}
    \end{subfigure}
    \hfill
    \begin{subfigure}{0.75\textwidth} 
        \centering
        \includegraphics[width=\textwidth]{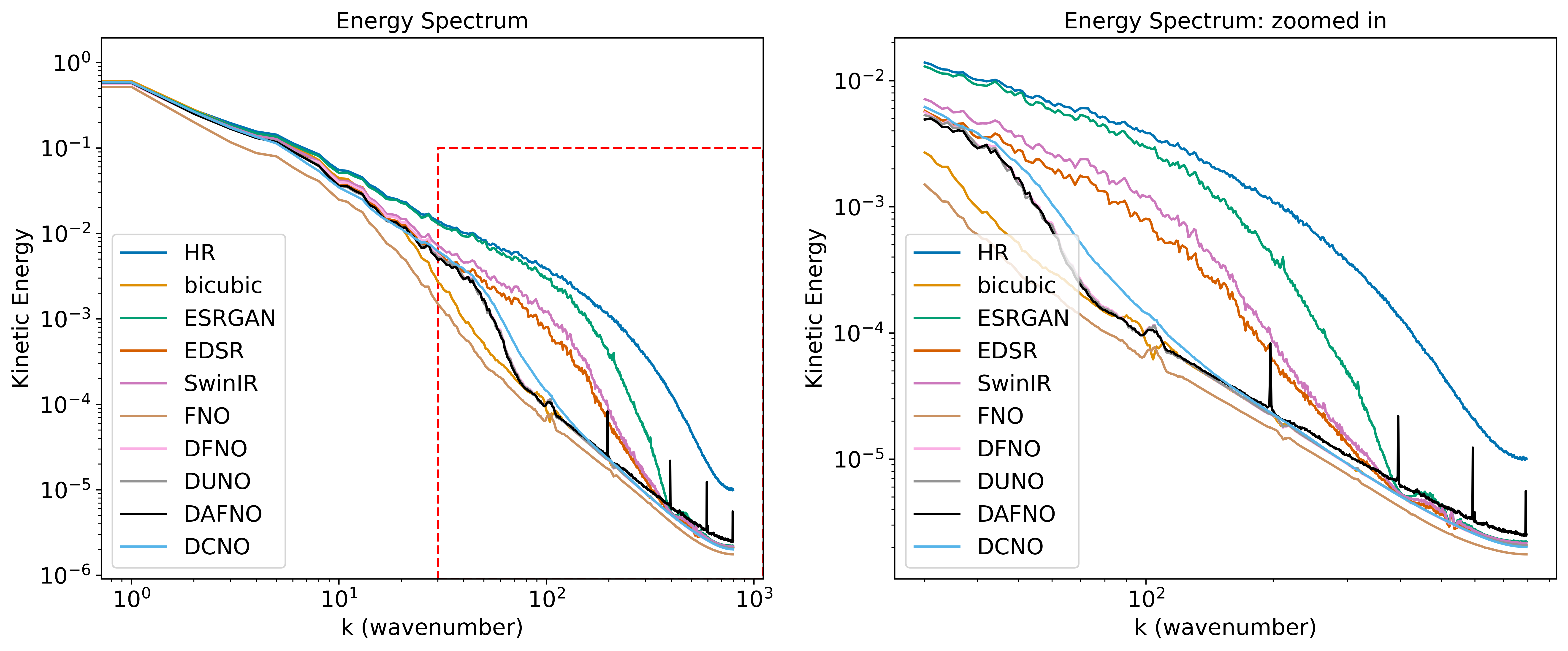} 
        \caption{}
    \label{fig:ke_wtk_zeroshot_spectrum}
    \end{subfigure}
    \caption{Figures (a) and (b) show kinetic energy spectrum plots for the ERA5 to WTK standard downscaling and zero-shot downscaling respectively. \textit{Kinetic Energy is normalized and wavenumber is measured relative to the domain size.} We add a zoomed-in plot (right) beside the main plot to zoom in on the key region of interest. ESRGAN matches the HR spectrum at all wavenumbers for
the standard downscaling (a). ESRGAN also comes closest to matching the ground truth energy spectrum for the
zero-shot downscaling (b), but the gap increases for higher wavenumbers. SwinIR and EDSR rank second to ESRGAN for both (a) and (b).
    }
    \label{fig:wtk_ke_plots}
\end{figure}

\begin{figure}[htbp]
    \centering
    \includegraphics[width=1\textwidth]{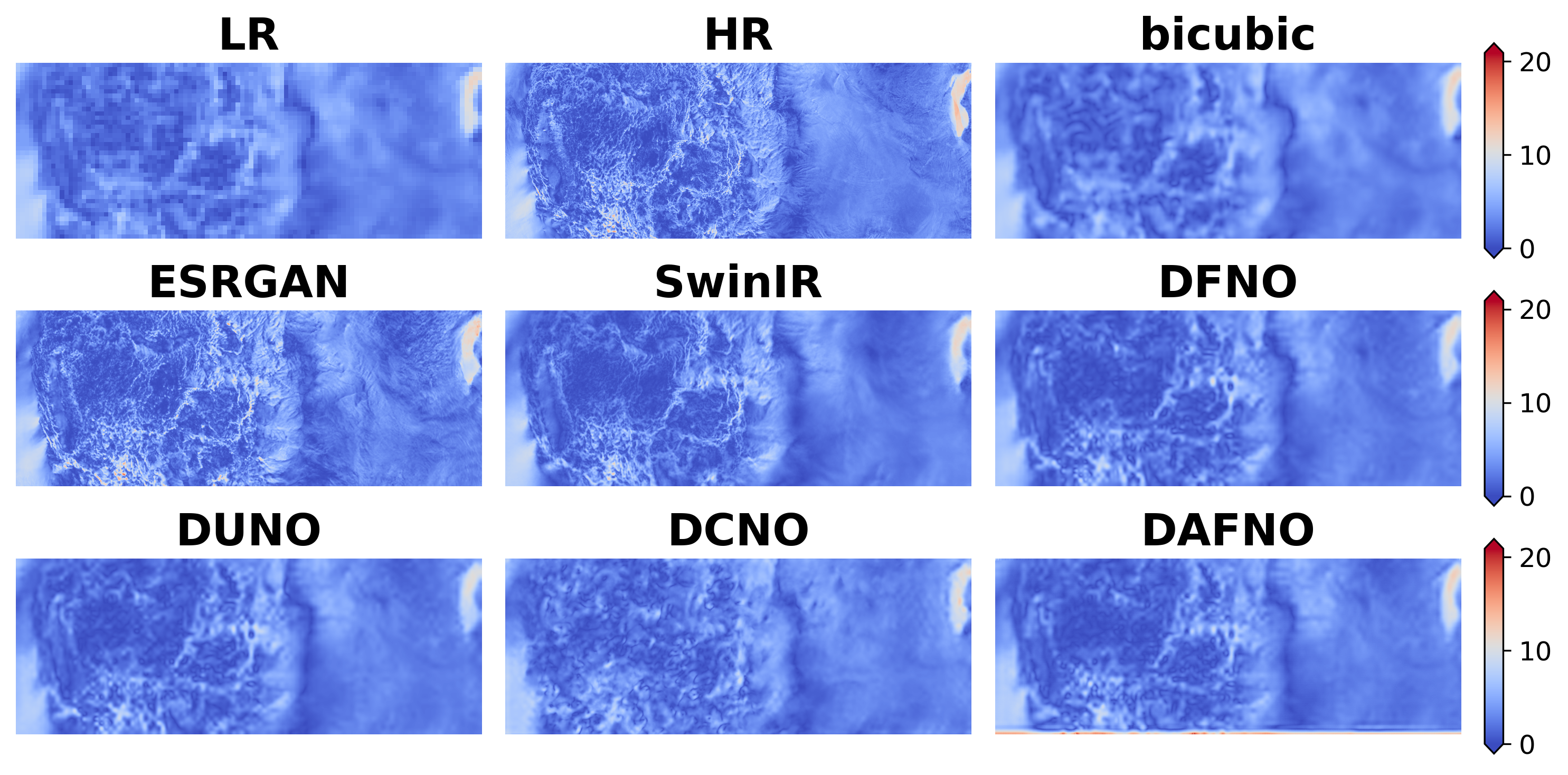}
    \caption{WTK wind speed visualization in $m/s$ generated from the zero-shot downscaling (figure shows results on one of the two regions). We observe ESRGAN's downscaled outputs (followed by SwinIR's) to be sharper with better details than the neural operator models.}
    \label{fig:viz_wtk_v_zeroshot}
\end{figure}

\textbf{Takeaways:} Overall, we observe the same relative model performance in this more difficult setting (ERA5 to WTK downscaling) as the easier setting (ERA5 to ERA5). The most significant takeaway remains that SwinIR is the best model at ERA5 to WTK zero-shot downscaling.
\\
\\
\textbf{Energy Spectrum} The energy spectrum plots for the ERA5 to WTK downscaling experiments are presented in Figures~\ref{fig:ke_wtk_spectrum} and ~\ref{fig:ke_wtk_zeroshot_spectrum}. ESRGAN matches
the HR spectrum at all wavenumbers for the standard downscaling. ESRGAN also comes closest to matching the HR energy spectrum for the zero-shot downscaling. However, as seen in Section~\ref{era_ds}, these models still underestimate the energy content in the high wavenumber range
for zero-shot downscaling. SwinIR and EDSR follow behind ESRGAN for both setups, but they outperform all the neural operator-based models. We observe DCNO to be close to the HR curve for most of the wavenumbers except for the highest ones for standard downscaling (Figure~\ref{fig:ke_wtk_spectrum}), but it is much worse at zero-shot downscaling (Figure~\ref{fig:ke_wtk_zeroshot_spectrum}). DAFNO no longer shows good performance for this experimental setup. FNO performs the worst, consistent with their performance in terms of the average error metrics (as seen in Table~\ref{table:wtk_5x}).
 Figure~\ref{fig:viz_wtk_v_zeroshot} compares the zero-shot downscaled outputs in terms of the wind speed for the bicubic interpolation, ESRGAN, SwinIR, DFNO, DUNO, DCNO, and DAFNO models. We see fine-scale features of the ground truth captured better in ESRGAN's downscaled outputs over other models.
 
\textbf{Takeaways:} ESRGAN is the closest to the energy spectrum of the ground truth HR data, but, there is an underestimation in the energy content for the higher wavenumber range, 
in the zero-shot downscaling. None of the neural operator models, including DAFNO, perform well in capturing the HR energy spectrum.




\subsection{Ablation study: Localized integral and differential kernel layers}
In this section, we ablate using localized operator layers~\cite{liu-schiaffini2024neural} in the FNO and FNO-based models.
It has recently been shown that global operations in FNOs limit their ability to capture local features well~\cite{liu-schiaffini2024neural}. Liu-Schiaffini et al.~\cite{liu-schiaffini2024neural} propose two localized operators: differential operators and integral operators with local kernels, and augment FNO layers with them to show improved performance on PDE learning tasks. We add these localized operators into the FNO and FNO-based models (DFNO and DUNO) and compare them against the same models without the local layers. 
Inspired by the strength of SwinIR and EDSR, we explore whether incorporating local layers might help the FNO-based models learn better local features and improve their performance in our downscaling task.

Table~\ref{table:ablation_study} shows the results of our ablation study on the effect of local layers~\cite{liu-schiaffini2024neural} on the FNO, DFNO, and DUNO modeling frameworks. Overall, the numbers do not indicate a significant impact on the MSE scores, especially for the ERA5 experiments. We observe FNO with the local layers to be better than without them on the ERA5 to WTK downscaling experiment, and DFNO and DUNO \emph{without the local layers} to be better across most experiments.
We suspect this might be pointing to the difference in architectures between the FNO and DFNO (or DUNO) models where the latter apply neural operators on the upsampled features obtained from RRDB layers. 
We note that the results in Sections~\ref{era_ds} and~\ref{era_wtk} are for FNO with local layers and the DFNO and DUNO results are without them.
\begin{table}[h]
    \centering
    \begin{tabular}{lccccc}
        \toprule
        & \textbf{Local layers} & \textbf{ERA5} & \textbf{ERA5 zero-shot} & \textbf{ERA5$\rightarrow$WTK} & \textbf{ERA5$\rightarrow$WTK zero-shot} \\
        \midrule
        FNO & \ding{55} & 0.83 & 0.74 & 7 & 7.5 \\
        FNO & \ding{51} & 0.91 & 0.71 & 5.69 & 5.45 \\
        DFNO & \ding{55} & 0.7 & 0.63 & 3.04 & 3.53 \\
        DFNO & \ding{51} & 0.66 & 0.66 & 4.63 & 5.31 \\
        DUNO & \ding{55} & 0.69 & 0.63 & 2.81 & 3.3 \\
        DUNO & \ding{51} & 0.62 & 0.67 & 3.03 & 3.79 \\
        \bottomrule
    \end{tabular}
    \caption{Ablation study on the effect of local layers from Liu-Schiaffini et al.~\cite{liu-schiaffini2024neural} on FNO, DFNO, and DUNO. We show the MSE in all the experiment setups. The local layers mostly benefit FNOs for the downscaling tasks but don't improve the performance of DFNO or DUNO models.}
    \label{table:ablation_study}
\end{table}
\section{Discussion}
In the literature, neural operators have performed well at zero-shot super-resolution~\cite{kovachki2023neural,DBLP:conf/iclr/LiKALBSA21,rahman2023uno,NEURIPS2023_f3c1951b} when trained to predict the solution of a PDE or when trained to act as an emulator of a time-dependent dynamical system.
Their resolution invariance property has also been utilized in Jiang et al.~\cite{jiang2023efficient} to train an FNO to act as an emulator for zero-shot super-resolution of weather forecasts. However, our results show that the neural operators under-perform the non-neural operators at zero-shot weather downscaling.
Importantly, physical system emulation differs from our static downscaling setting, where we train on pairs of low and high-resolution images. 
In our case, the neural operators are trained to learn a mapping between resolutions, and are tested on their ability to generalize zero-shot to higher upsampling factors.
We believe this distinction is important to help contextualize our results, which show that neural operators have difficulty with this task.

Our results show that bicubic interpolation followed by a vanilla FNO performs poorly at weather downscaling. In most cases, this performs worse than bicubic interpolation alone. FNO learns spectral features in Fourier space which makes it resolution-invariant---these features are not inherently tied to the resolution of the training dataset.  
Weather downscaling may benefit from learning spatial features tied to the specific input \textit{and} output grid resolutions. This could be limiting the vanilla FNO's ability to downscale well. We also evaluate neural operator models with convolutional RRDB blocks before the neural operator layers (the D\textit{X}NO models (Section~\ref{methodology_experiments})), which improves the downscaling performance significantly. 
The DCNO models, based on CNOs~\cite{NEURIPS2023_f3c1951b}, adapt U-Net style convolutions to approximately learn an operator mapping. They perform close to the best model SwinIR in the standard downscaling experiments, but, their performance drops significantly in the zero-shot setup, as CNO uses explicit up/down-sampling and thus cannot be applied to different resolutions without some degraded performance~\cite{liu-schiaffini2024neural}. 

ESRGAN proves to be the best model for capturing the physical properties of the data at medium-to-high wavenumbers for ERA5 to ERA5 and all wavenumbers for ERA5 to WTK experiments, as measured in our work by kinetic energy plots, for zero-shot downscaling. It is important to note that zero-shot downscaling is a challenging task as we expect the models to produce outputs that have super-resolved physics at the finer scale without training on them. It is possible that ESRGAN learns to generate downscaled outputs with better visual quality because of its architectural design and use of perceptual loss, which may help in capturing the HR physics across spatial scales, yet, we observe that all models underestimate the energy content in the high-wavenumber range for zero-shot downscaling. It seems that SwinIR learns superior-quality features at the smaller upsampling factor during training, enabling an interpolation on top of SwinIR to generate downscaled outputs better than other models as shown by the average error metrics. We did an ablation study where we replaced the convolutional RRDB blocks in D\textit{X}NO models with the residual Swin-Transformer blocks (RSTB) as adopted in SwinIR to compare their downscaling performances (details in Appendix~\ref{ablation_study_rstb}).
D\textit{X}NO models with SwinIR-based feature extraction performed worse than D\textit{X}NO with RRDB modules, suggesting more advanced hybrid neural-operator-transformer architecture~\cite{luo2024hierarchical} may be needed. We recommend that researchers benchmark against powerful non-operator-learning methods with interpolation as strong baselines. However, given that SwinIR and ESRGAN need to use bicubic interpolation (which has no learnable parameters) to do zero-shot downscaling, it could be fundamentally limited in its ability to downscale small-scale physics unseen during training. It is also possible that the set of neural operator models we explored can be improved.
Overall, all models appear to be quite far from solving our downscaling tasks. 



\section{Conclusion}
This work comprehensively benchmarks neural operators on the task of weather downscaling, with a particular emphasis on critically investigating the zero-shot downscaling capabilities of neural operators. 
Our analyses involve two studies over (1) learning a mapping from coarsened ERA5 to high-resolution ERA5, and (2) learning a mapping from low-resolution ERA5 wind data to a high-resolution wind data (2km x 2km). 
Our zero-shot downscaling experiments involve challenging upsampling factors: 8x and 15x over the two studies respectively. 

With an extensive evaluation using various error metrics and kinetic energy spectrum plots, we show that resolution-invariant neural operators are outperformed by the Swin-Transformer and ESRGAN-based models, even at zero-shot downscaling. This was surprising, as resolution-invariant neural operators were previously shown to be good at zero-shot super-resolution for the task of emulating dynamical systems. While our current study presents limitations of neural operators at weather downscaling, future research may consider improving the neural operator downscaling frameworks with better feature encoders or advanced hybrid neural-operator-transformer models. We hope this work provides a deeper understanding of the role of neural operators in weather downscaling and furthers research in this direction.

\paragraph{Acknowledgments}

This work was authored by the National Renewable Energy Laboratory (NREL), operated by Alliance for Sustainable Energy, LLC, for the U.S. Department of Energy (DOE) under Contract No. DE-AC36-08GO28308. This work was supported by the Laboratory Directed Research and Development (LDRD) Program at NREL. The views expressed in the article do not necessarily represent the views of the DOE or the U.S. Government. The U.S. Government retains and the publisher, by accepting the article for publication, acknowledges that the U.S. Government retains a nonexclusive, paid-up, irrevocable, worldwide license to publish or reproduce the published form of this work, or allow others to do so, for U.S. Government purposes. 
The research was performed using computational resources sponsored by the Department of Energy's Office of Energy Efficiency and Renewable Energy and located at the National Renewable Energy Laboratory.





\bibliographystyle{unsrt}
\bibliography{main}

\newpage
\appendix

\section{Training details}
\subsection{Additional data details}
\label{data_details}
\begin{enumerate}
\item \textbf{ERA5 to ERA5 downscaling}: 
The number of samples (image snapshots) in training, validation, and test are 1460, 730, and 730 respectively. Refer to the main text for the size details (height, width, channels) of each image.\\ 
\\
We train all the models two times in this setup. \\
\textit{Standard downscaling}: This involves training all the models with an upsampling factor of 8x. Following Ren et al.~\cite{ren2023superbench}, we do not use the entire image snapshots but random crops of the images for training. We extract eight patches of size 128x128 from each image snapshot to obtain HR image data for training. The corresponding LR image patches of size 16x16 are created by coarsening the HR patches with bicubic interpolation. \\
\textit{Zero-shot downscaling}: This involves training all the models with an upsampling factor of 4x. For this case, we extract eight patches of size 64x64 from each image snapshot to obtain HR image data for training. The corresponding LR image patches of size 16x16 are created by coarsening the HR patches with bicubic interpolation.
\item \textbf{ERA5 to WTK downscaling}:
We work with two regions in the US. For each of the regions: the number of samples (image snapshots) in training, validation, and test are 7008, 1752, and 8760 respectively. Refer to the main text for the size details (height, width, channels) of each image.\\

All the models are trained just once in this setup, with an upsampling factor of 5x. We again use random crops of the images for training. We extract eight patches of size 160x160 from each image snapshot to obtain HR (WTK) image data for training. The LR image patches of size 32x32 are obtained from the corresponding ERA5 image snapshots. We train a single model for both regions, ensuring that every training batch has an
equal number of LR and HR pairs from both regions. 
\end{enumerate}
The crop size for the ERA5 to ERA5 experiments is the same as the one used in SuperBench.~\cite{ren2023superbench}. We tune the crop size for the ERA5 to WTK downscaling experiments and obtain the optimal crop size reported above.

\subsection{Hyperparameters}
\label{hyperparameter_details}
All the models are trained for 400 epochs. \\
\\
\textbf{Neural-operator-based models} : \\
All the hyperparameters are tuned over the validation dataset. We perform a sweep over learning rates \{0.005,0.0001,0.00001\} for all models. The models are trained using the ADAM~\cite{DBLP:journals/corr/KingmaB14} optimizer with a batch size of 32, weight decay of 1e-4, and a step learning rate scheduler with a step size of 60. For the Downscaling (D) models, we add the RRDB module (a component of the ESRGAN framework~\cite{wang2018esrgan}), before the bicubic interpolation layer and the subsequent neural operator layers. RRDB implementation is adapted from the ESRGAN implementation discussed below. We perform a sweep over the number of RRDB blocks \{6,12,24\} for all the downscaling models.
\begin{enumerate}
\item \textbf{FNO} We follow the original implementation of FNO from the \textit{neuraloperator} library~\cite{kossaifi2024neural}, based on~\cite{DBLP:conf/iclr/LiKALBSA21,kovachki2023neural}, incorporating localized operator layers from~\cite{liu-schiaffini2024neural} and using most of the default model hyperparameters. We perform a hyperparameter sweep over the number of hidden channels in the lifting and projection blocks, testing values \{128,256\}, selecting 256 for them. We also examine the number of modes to keep in the Fourier layers - extensive study detailed in~\ref{fno_mode_analysis}, selecting the optimal number of modes as 16 for the ERA5 to WTK experiments and 8 for ERA5 to ERA5 experiments. The best learning rate for the FNO model is found to be 0.005. We use the $lp$ loss with $p=2$, reduced over $dim=0$ as defined in the original implementation. 
\item \textbf{DFNO}  We keep the selected values for the lifting, projection channels, and the number of modes, obtained from tuning the above FNO model. The best learning rate for training DFNO is found to be 0.0001, and the optimal number of RRDB blocks is selected as 12. MSE is used as the loss function. 
\item \textbf{DUNO} We follow the UNO model implementation, again from the \textit{neuraloperator} library~\cite{kossaifi2024neural,kovachki2023neural}. While we use most of the default model hyperparameters, we do hyperparameter tuning on the hidden channels (initial width of UNO) over \{32,64\}, selecting 64 as optimal, and the number of output channels of each Fourier layer over \{32,64\}, selecting them as 64. We found the best learning rate to be 0.0001, and the optimal number of RRDB blocks is 12. MSE is used as the loss function.
\item \textbf{DAFNO} We follow the implementation of the AFNO network from the FourCastNet~\cite{pathak2022fourcastnet} implementation. With most model hyperparameters as default, we perform a hyperparameter sweep over the patch size \{4,8\}, choosing 8 as optimal, and the number of blocks (block as defined in~\cite{guibas2021adaptive}) \{4,8\}, selecting 8. It should be noted that the optimal patch size is found to be 4 when training for the ERA5 to ERA5 zero-shot downscaling setup. For the DAFNO training, we find the best values for the learning rate to be 0.0001, and the number of RRDB blocks to be 12. MSE is used as the loss function.
\item \textbf{DCNO} We follow the original CNO implementation from Raonic et al.~\cite{NEURIPS2023_f3c1951b} with most of the the default model hyperparameters. We tune the number of layers (upsampling/downsampling blocks) over \{3,4\} and find the optimal to be 3. For the DCNO training, we find the best values for the learning rate to be 0.0001, and the number of RRDB blocks to be 12. MSE is used as the loss function.
\\
\end{enumerate}
\textbf{Baseline models} : \\
The implementations of SRCNN, EDSR, and SwinIR model pipelines follow the implementations provided by Ren et al.~\cite{ren2023superbench}. We follow an open-source implementation of ESRGAN from Li~\cite{ESRGANPyTorch} for our ESRGAN downscaling framework. We keep intact most of the hyperparameters and training setups from these implementations, but we train each of the baseline models for a fixed 400 epochs (consistent with the neural-operator-based models). We also do a hyperparameter sweep over the learning rates \{0.001,0.0001,0.00001\} for all the baseline models, using the validation dataset to tune this hyperparameter. For the ERA5 to ERA5 experiments: we find the optimal learning rate as 0.0001 for EDSR, SwinIR, and ESRGAN, and 0.001 for SRCNN. For the ERA5 to WTK experiments: we find the optimal learning rate as 0.0001 for SRCNN, SwinIR, and ESRGAN, and 0.001 for EDSR. 

\subsection{Model parameters and Training wall-clock time}
\label{parameter_details}
\begin{table}[h]
\centering
\caption{Model parameters and training wall-clock time recorded on a single NVIDIA H100 GPU for all the baseline and neural-operator-based models used in the ERA5 to WTK downscaling setup. 
SwinIR achieves superior average error metrics (e.g., MSE), as shown in Table~\ref{table:wtk_5x}, while having only marginally higher model parameters than the downscaling neural-operator-based models (except DAFNO). ESRGAN is the best model in matching the ground truth energy spectrum (Figures~\ref{fig:ke_wtk_spectrum},~\ref{fig:ke_wtk_zeroshot_spectrum}) and is second to DAFNO in parameter count. Notably, ESRGAN takes the longest to run, followed by SwinIR; both take longer to run than all the neural-operator-based models.
}
\label{tab:example}
\begin{tabular}{lll} 
\toprule
\textbf{Model} & \textbf{\#Parameters} & \textbf{Training time (hours)} \\ 
\midrule
SRCNN & 0.063M & 0.52 \\ 
ESRGAN & 39.18M & 14.84 \\ 
EDSR & 2.14M & 0.57 \\ 
SwinIR & 12.53M & 10.77 \\ 
\midrule
FNO & 1.24M & 2.86 \\ 
DFNO & 9.88M & 4.8 \\ 
DUNO & 9.36M & 6.73 \\ 
DAFNO & 69.15M & 7.64 \\ 
DCNO & 11.33M & 4.83 \\ 
\bottomrule
\end{tabular}
\end{table}

\subsection{Analysis of FNO frequency-cutoff}
\label{fno_mode_analysis}
We conduct a study to investigate the impact of the frequency cutoff, or the number of modes retained in the Fourier layers~\cite{DBLP:conf/iclr/LiKALBSA21,kovachki2023neural}, on the downscaling performance of FNO. This hyperparameter specifies the number of frequency components retained for training after a Fourier transform is applied to the data. As with other hyperparameters, we select the number of modes for the FNO model that yields the lowest error on the validation dataset.
\begin{enumerate}
 \item \textbf{ERA5 to WTK downscaling}:
    We need to train a single FNO model for this setup. With image crops of size 160x160 used in training, we perform a sweep over \{16, 32, 64, 128, 160\} number of modes and observe the lowest train, validation and test MSE with 16 modes. Figure~\ref{fig:mse_fno_wtk} shows the effect of the number of modes on the test MSE for both the standard downscaling and zero-shot downscaling. We also show the train MSE and validation MSE (for the standard downscaling task) in this figure. Thus, we select 16 as the optimal number of modes for this setting. Additionally, Figure~\ref{fig:fno_modes_energy_spectra_wtk} presents energy spectrum plots comparing the number of modes for downscaling on the train, validation, and test datasets. We again note that FNO with 16 modes is better in capturing the ground truth energy spectra, even at high wavenumbers, though the energy content underestimation becomes more pronounced with higher wavenumbers.
    \begin{figure}[htbp]
    \centering
    \includegraphics[width=0.6\textwidth]{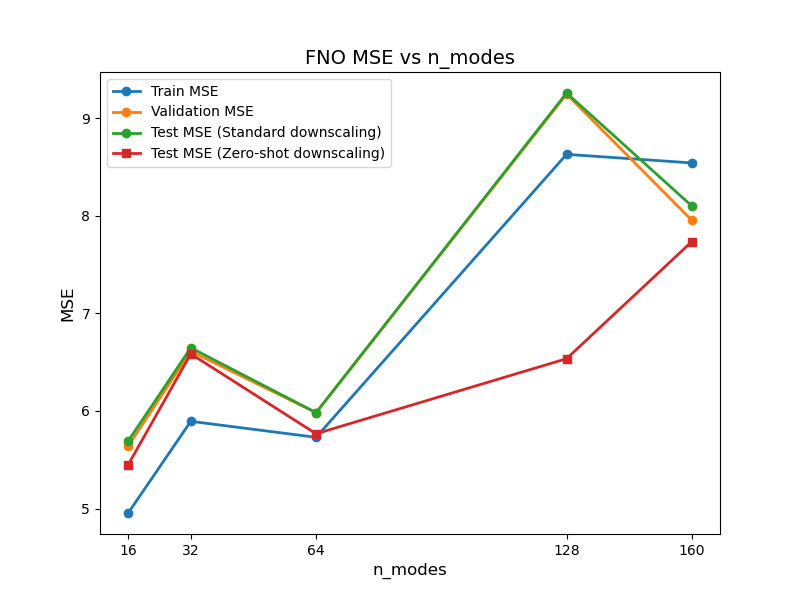}
    \caption{
    Plot comparing the mean-squared error (MSE) against the number of modes (n\_\text{modes}) in the Fourier Neural Operator (FNO) model for ERA5 to WTK downscaling. 
    }
    \label{fig:mse_fno_wtk}
\end{figure}
\begin{figure}[htbp]
    \centering
    \begin{subfigure}[t]{0.45\textwidth} 
        \centering
        \includegraphics[width=\textwidth]{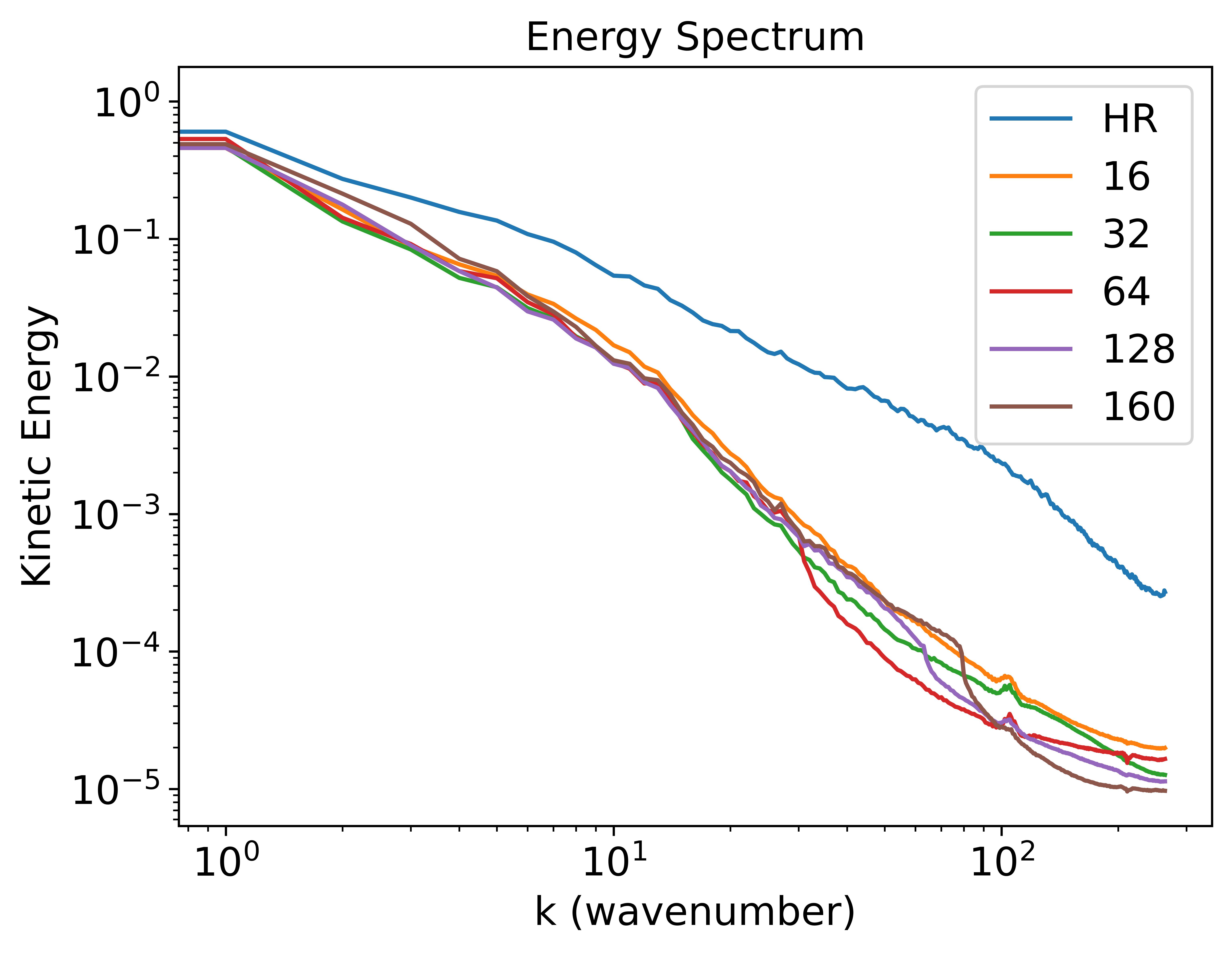}
        \caption{Energy spectrum plot on the train data.}
    \end{subfigure}
    \hfill
    \begin{subfigure}[t]{0.45\textwidth} 
        \centering
        \includegraphics[width=\textwidth]{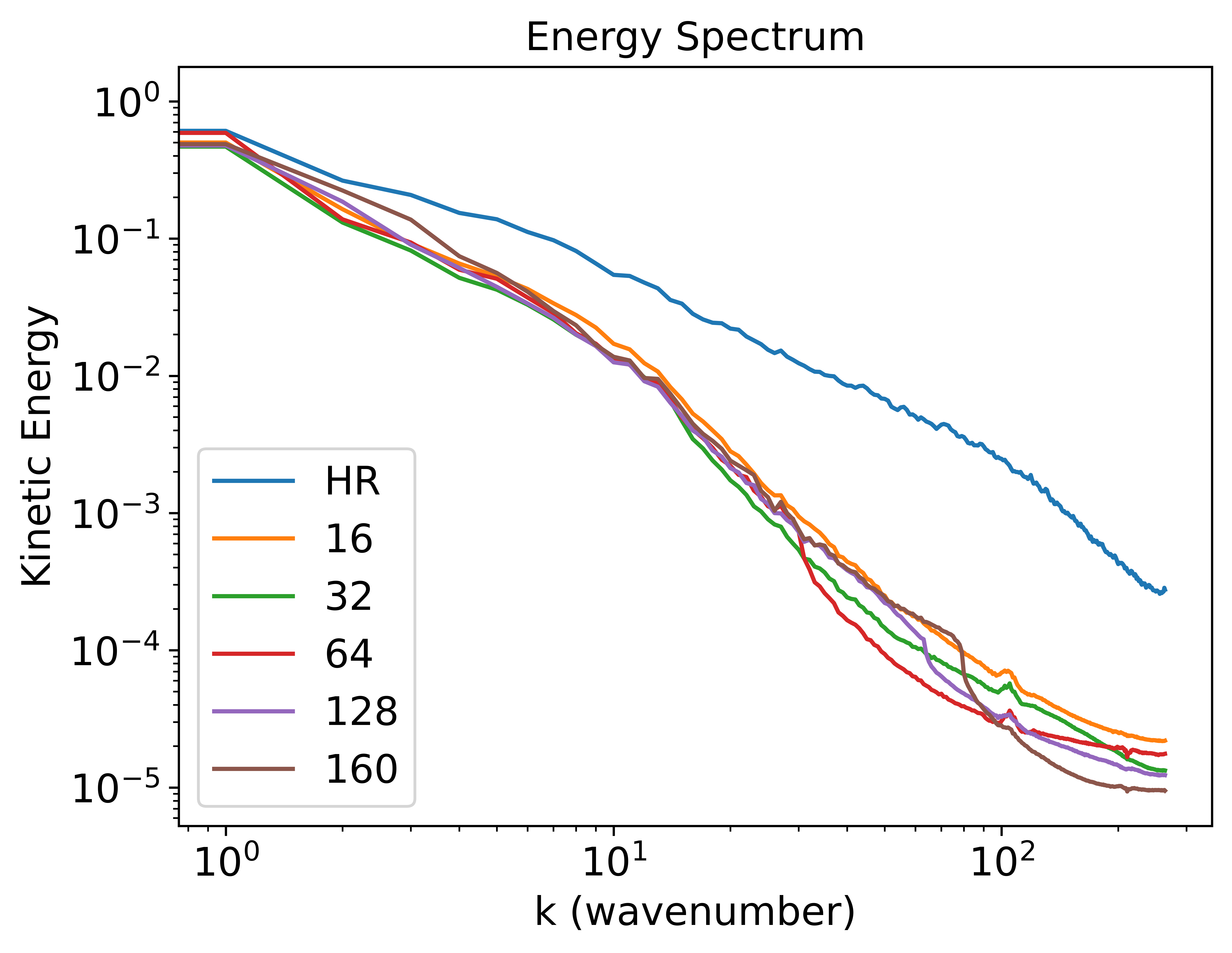}
        \caption{Energy spectrum plot on the validation data.}
    \end{subfigure}


    \begin{subfigure}[t]{0.45\textwidth} 
        \centering
        \includegraphics[width=\textwidth]{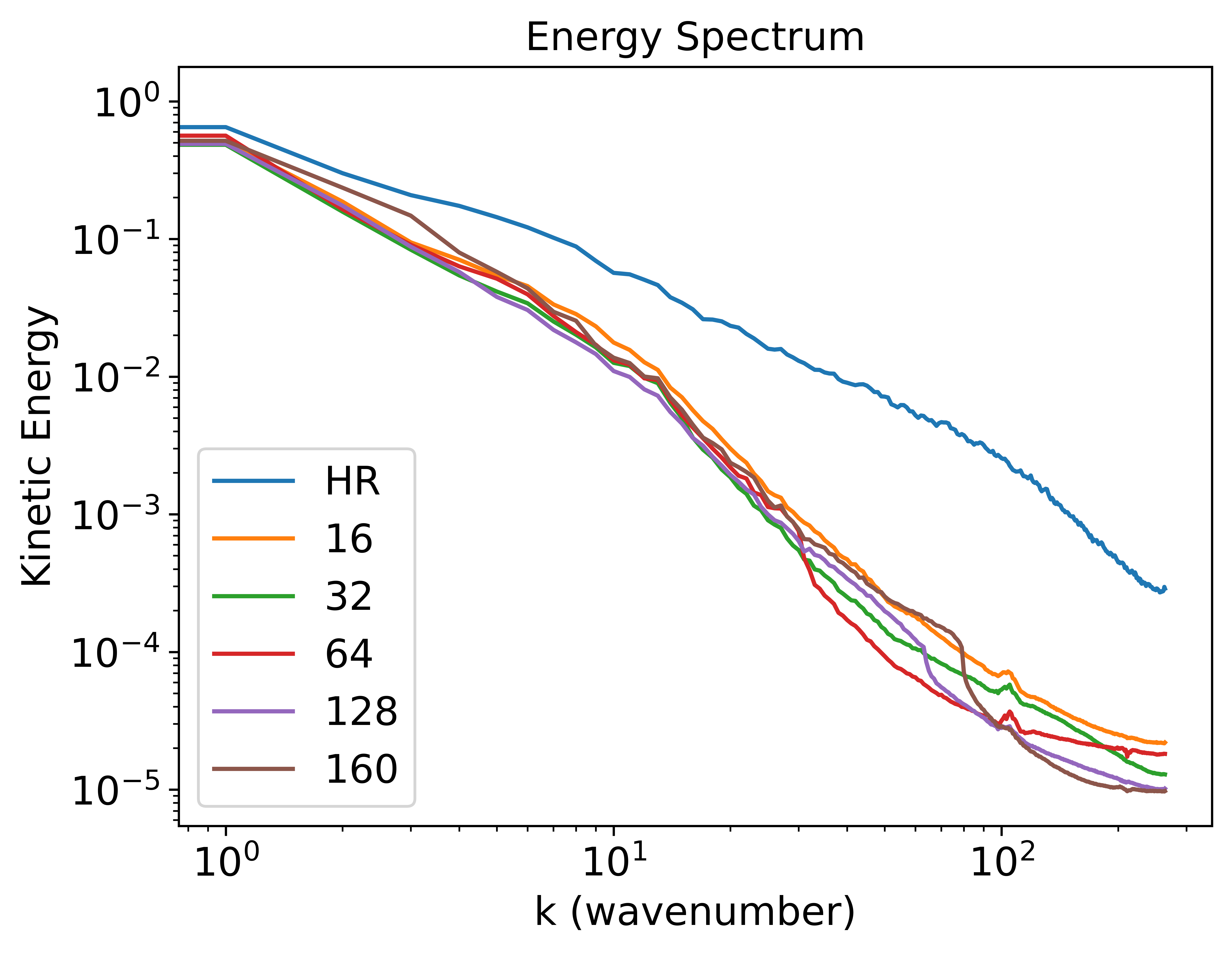}
        \caption{Energy spectrum plot on the test data (standard downscaling task).}
    \end{subfigure}
    \hfill
    \begin{subfigure}[t]{0.45\textwidth} 
        \centering
        \includegraphics[width=\textwidth]{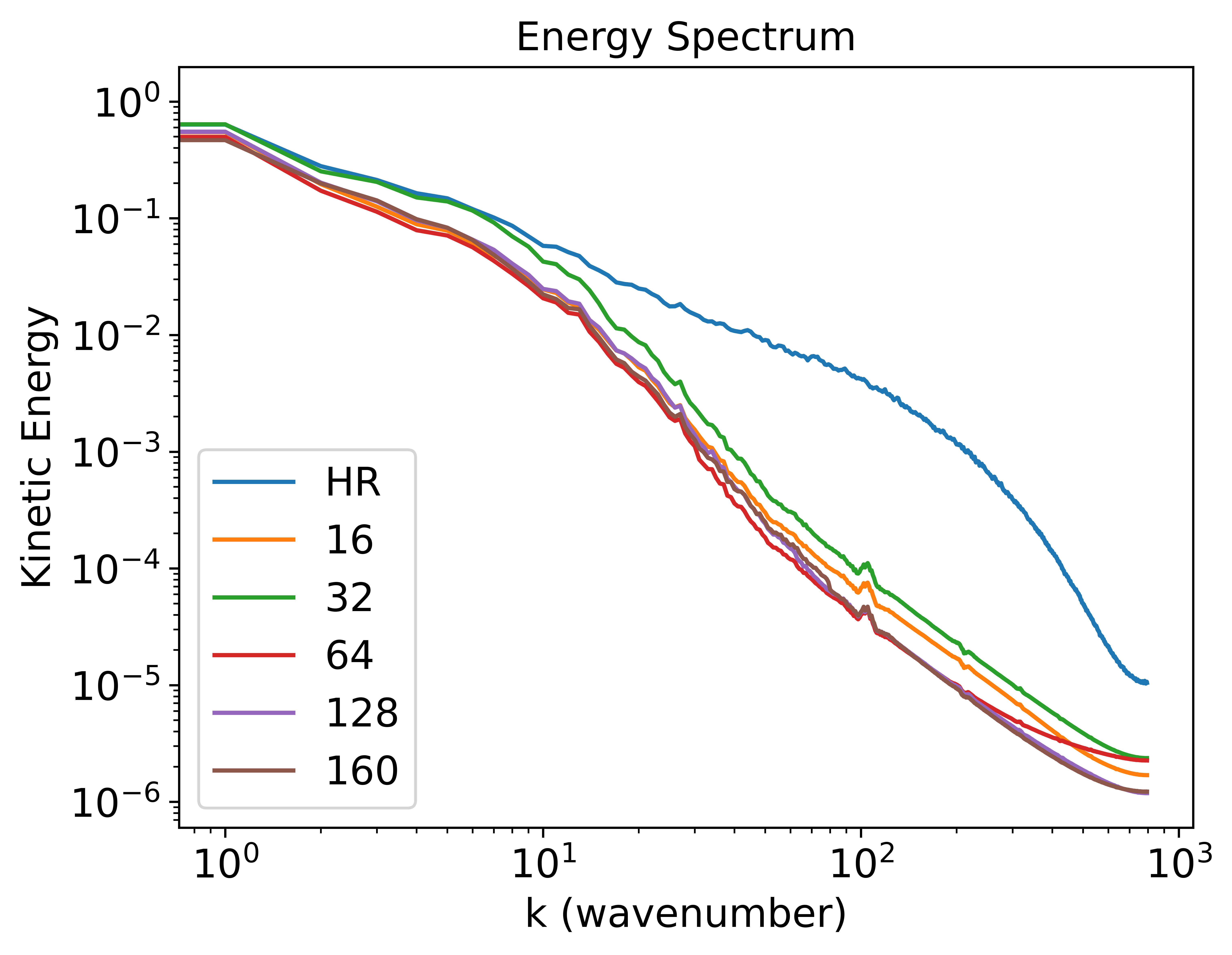}
        \caption{Energy spectrum plot on the test data (zero-shot downscaling task).}
    \end{subfigure}

    \caption{Energy spectrum plots comparing the effect of different numbers of modes (\{16, 32, 64, 128, 160\}) in the FNO model for ERA5 to WTK downscaling.}
    \label{fig:fno_modes_energy_spectra_wtk}
\end{figure}
    \item \textbf{ERA5 to ERA5 downscaling}:
    The FNO model is trained twice in this setup: \\
    \textit{Standard downscaling: } With image crops of size 128x128 used in training the FNO model in this case, we perform a sweep over \{8, 16, 32, 64, 128\} number of modes and observe the lowest train, validation and test MSE with 8 modes. Thus, we select 8 as the optimal number of modes for this setting.  \\ 
    \textit{Zero-shot downscaling: } With image crops of size 64x64 used in training the FNO in this case, we perform a sweep over \{8, 16, 32, 64\} number of modes and observe the lowest train, validation and test MSE with 8 modes. Thus, we select 8 as the optimal number of modes for this setting. 

    Figure~\ref{fig:mse_fno_era} shows the effect of the number of modes on the test MSE for both the standard downscaling and zero-shot downscaling. The train and validation MSE scores show similar trends. The energy spectra plots for ERA5 to ERA5 downscaling do not provide any additional insights.
    \begin{figure}[htbp]
    \centering
    \includegraphics[width=0.6\textwidth]{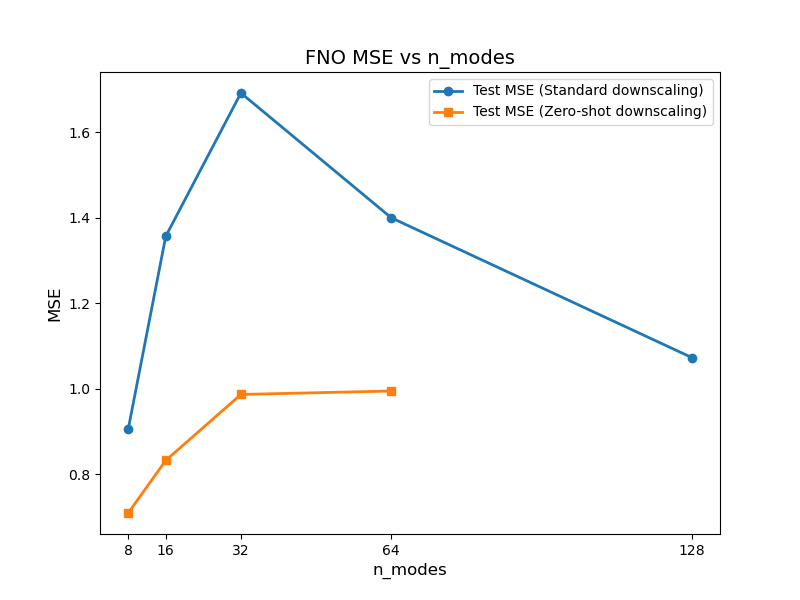}
    \caption{
    Plot comparing the MSE against the number of modes (n\_\text{modes}) in the FNO model for ERA5 to ERA5 downscaling. In this case (as discussed in~\ref{fno_mode_analysis}), the standard downscaling FNO model performs a sweep over \{8, 16, 32, 64, 128\} number of modes, and the zero-shot downscaling model sweeps over \{8, 16, 32, 64\}
number of modes.
    }
    \label{fig:mse_fno_era}
\end{figure}
\end{enumerate}

Overall, we observe that the train, validation, and test MSE are lowest when using the smallest number of modes across all setups. Moreover, the energy spectra plots for the lowest mode match more closely with the HR energy spectra, capturing the physics of the ground truth data better. 
While further investigation is needed to understand our observation better, these findings provide useful insights into the effect of FNO frequency-cutoff on its performance.

\section{Ablation study on D\textit{X}NO models}
\label{ablation_study_rstb}
Our superior results with SwinIR suggest that residual Swin-Transformer blocks (RSTB), as adopted in SwinIR, are good at extracting high-quality features for the task of downscaling. This motivated us to explore whether they could enhance feature extraction compared to the convolutional RRDB blocks when used in the D\textit{X}NO models in our experiments. We replaced the RRDB blocks in the D\textit{X}NO model with this transformer-based feature extraction module from SwinIR, keeping the other parts of the D\textit{X}NO model (the interpolation and neural operator layers) unchanged. Table~\ref{table:ablation_study_rstb} shows that simply swapping the RRDB blocks with the RSTB blocks worsens the downscaling model's performance. It is possible that SwinIR's effectiveness can be attributed to its overall architecture, including the upsampling module and other layers that complement the RSTB feature extraction better. We may also need advanced models to integrate the potential advantages of transformers with neural operators for the task of downscaling. The table includes the ablation study results for the ERA5 to WTK standard and zero-shot downscaling experiments. Notably, the DFNO model with RSTB blocks failed to converge when trained for the ERA5 to ERA5 downscaling setup. 
\begin{table}[htbp]
    \centering
    \begin{tabular}{lccccc}
        \toprule
        & \textbf{Feature extractor} &  \textbf{ERA5$\rightarrow$WTK} & \textbf{ERA5$\rightarrow$WTK zero-shot} \\
        \midrule
        DFNO & RRDB & 3.04 & 3.53 \\
        DFNO & RSTB & 5.69 & 6.19 \\
        DUNO & RRDB & 2.81 & 3.3 \\
        DUNO & RSTB & 4.65 & 5.15  \\
        \bottomrule
    \end{tabular}
    \caption{Ablation study on the effect of replacing convolutional RRDB blocks with residual Swin-Transformer blocks (RSTB) in the D\textit{X}NO models. The table shows the MSE for the ERA5 to WTK downscaling experiments using the DFNO and DUNO models. D\textit{X}NO with RRDB blocks show better performance.}
    \label{table:ablation_study_rstb}
\end{table}
\section{Additional ERA5 to ERA5 downscaling results}
\label{appendix_era}
\begin{table}[htbp]
    \centering
    \caption{\textbf{ERA5 to ERA5 temperature downscaling results}. MSE has units $(K)^2$, MAE $K$ and IN $K$.  We bold the best-performing model among all the models and underline the best-performing neural operator model.
    \label{table:era5_temp_8x} }
    \begin{adjustbox}{max width=\columnwidth}
    \begin{tabular}{lccccccccc}
        \toprule
        & & \multicolumn{4}{c}{\textbf{Standard Downscaling}}           & \multicolumn{4}{c}{\textbf{Zero-shot Downscaling}}           \\ \cmidrule(l){2-10}
        & is NO? & MSE $\downarrow$ & MAE $\downarrow$ & IN $\downarrow$ & PSNR $\uparrow$ & MSE $\downarrow$ & MAE $\downarrow$ & IN $\downarrow$ & PSNR $\uparrow$ \\
        \midrule
        bicubic & \ding{55}  & 2.47  & 0.89 & 20.41  & 45.99   & 2.47  & 0.89 & 20.41  & 45.99     \\
        SRCNN & \ding{55} & 2.09  & 0.83  & 20.94  & 46.74   & 1.95  & 0.78  & 20.35 & 47.02   \\
        ESRGAN & \ding{55} & 3.75  & 1.27  & 45.5  & 44.18   & 1.54  & 0.82  & 16.88  & 48.04   \\
        EDSR & \ding{55} & 0.68  & 0.44 & 15.5 & 51.6  & 0.81  & 0.47 & 15.91 & 50.84     \\
        SwinIR & \ding{55} & \textbf{0.38} & \textbf{0.34}  & \textbf{11.09}  & \textbf{54.15}  & \textbf{0.81} & \textbf{0.46}  & \textbf{15.88}  & \textbf{50.84}          \\
        \midrule
        FNO & \ding{51} & 1.12  & 0.6  & 17.68 & 49.41  & 1.08  & 0.57  & 17.84  & 49.59  \\
        DFNO & \ding{51} & 0.94  & 0.55  & 16.38  & 50.17  & 0.95  & 0.54  & 17.14  & 50.13  \\
        DUNO & \ding{51} & 0.88  & 0.52 & 15.61 & 50.47  & \underline{0.94}  & \underline{0.53} & \underline{17.18} & \underline{50.2}   \\
        DAFNO & \ding{51} & 0.9  & 0.56  & 15.45  & 50.39 & 1.21  & 0.68  & 32.75  & 49.1     \\
        DCNO & \ding{51} & \underline{0.45}  & \underline{0.4}  & \underline{11.11} & \underline{53.38}  & 1.85  & 0.81  & 16.72 & 47.26    \\
        \bottomrule
    \end{tabular}
    \end{adjustbox}
\end{table}

\begin{table}[htbp]
    \centering
    \caption{\textbf{ERA5 to ERA5 total column water vapor downscaling results}. MSE has units $(kg/m^2)^2$, MAE $kg/m^2$ and IN $kg/m^2$.  We bold the best-performing model among all the models and underline the best-performing neural operator model.}
    \label{table:era_8x_tchw}
    \begin{adjustbox}{max width=\columnwidth}
    \begin{tabular}{lccccccccc}
        \toprule
          & & \multicolumn{4}{c}{\textbf{Standard Downscaling}}           & \multicolumn{4}{c}{\textbf{Zero-shot Downscaling}}           \\ \cmidrule(l){2-10}
        & is NO? & MSE $\downarrow$ & MAE $\downarrow$ & IN $\downarrow$ & PSNR $\uparrow$ & MSE $\downarrow$ & MAE $\downarrow$ & IN $\downarrow$ & PSNR $\uparrow$ \\
        \midrule
        bicubic & \ding{55}  & 2.72  & 0.98 & 29.44  & 33.18   & 2.72  & 0.98 & 29.44  & 33.18   \\
        SRCNN & \ding{55} & 2.59  & 0.96  & 29.24  & 33.4 & 2.27  & 0.89  & 29.01  & 33.97    \\
        ESRGAN & \ding{55} & 3.1  & 0.94  & 48.3  &  32.61  & 1.71  & 0.79  & 26.17  &  35.19  \\
        EDSR & \ding{55} & 0.94  & 0.55 & 23.76 & 37.81   & 2.92  & 1.05 & 15.61 & 21.2     \\
        SwinIR & \ding{55} & \textbf{0.61} & \textbf{0.46}  & \textbf{16.07}  & \textbf{39.68}  & \textbf{1.09} & \textbf{0.56}  & \textbf{26.72}  & \textbf{37.15}      \\
        \midrule
        FNO & \ding{51} & 1.88  & 0.79  & 29.31  & 34.77 & 1.67 & 0.76  & 27.74 & 35.28    \\
        DFNO & \ding{51} & 1.21  & 0.65 & 25.29  & 36.68   & 1.38 & 0.66  & 26.03  & 36.14  \\
        DUNO & \ding{51} & 1.18  & 0.64 & 25.47 & 36.81  &  \underline{1.38}  &  \underline{0.66} &  \underline{26.14} &  \underline{36.13}  \\
        DAFNO & \ding{51} & 1.16  & 0.65  & 25.28  & 36.87  & 1.85  & 0.84 & 26.62  & 34.84   \\
        DCNO & \ding{51} &  \underline{0.67}  &  \underline{0.51}  &  \underline{13.53} &  \underline{39.23}  & 2.53  & 1.01 & 28.45 & 33.48   \\
        \bottomrule
    \end{tabular}
    \end{adjustbox}
\end{table}


\end{document}